\documentclass[aps,prl,nofootinbib,twocolumn,showpacs,superscriptaddress,letterpaper,amsmath,amssymb]{revtex4}
\usepackage{graphicx}  
\usepackage{dcolumn}   
\usepackage{bm}        
\usepackage{amssymb}   
\usepackage{feynmf}
\usepackage{slashed}
\usepackage{multirow}
\usepackage{color}
\usepackage{array}
\usepackage{url}
\usepackage{tikz}
\usepackage[normalem]{ulem}

\definecolor{cadmiumgreen}{rgb}{0.0, 0.42, 0.24}
\newcommand{\CNNresult}[0]{0.841}
\newcommand{\PFNresult}[0]{0.857}
\newcommand{\ISOOneresult}[0]{0.787}
\newcommand{\ISOTworesult}[0]{0.793}
\newcommand{\ISOTenresult}[0]{0.803}

\newcommand{\rev}[1]{#1}

\newcommand{\rvv}[1]{#1}
\newcommand{\rvvv}[1]{ #1}

\newcommand{\showefp}[3]{
\node at (a.north east)
    [
    anchor=center,
    xshift=-42mm,
    yshift=-12mm
    ]
    {
    \setlength{\fboxrule}{0.02pt}%
    $\gPlot{0.035}{#1}$
    };
    \node at (a.north east)
    [
    anchor=center,
    xshift=-40mm,
    yshift=-17mm
    ]
    {
    \setlength{\fboxrule}{0.01pt}%
    \scalebox{0.6}{$\kappa=#2,\beta=#3$}
    };
    }
    
    \newcommand{\showefpR}[3]{
\node at (a.north east)
    [
    anchor=center,
    xshift=-12mm,
    yshift=-12mm
    ]
    {
    \setlength{\fboxrule}{0.02pt}%
    $\gPlot{0.035}{#1}$
    };
    \node at (a.north east)
    [
    anchor=center,
    xshift=-12mm,
    yshift=-17mm
    ]
    {
    \setlength{\fboxrule}{0.01pt}%
    \scalebox{0.6}{$\kappa=#2,\beta=#3$}
    };
    }
    
\newcommand{\showefpRk}[2]{
\node at (a.north east)
    [
    anchor=center,
    xshift=-12mm,
    yshift=-12mm
    ]
    {
    \setlength{\fboxrule}{0.02pt}%
    $\gPlot{0.035}{#1}$
    };
    \node at (a.north east)
    [
    anchor=center,
    xshift=-12mm,
    yshift=-17mm
    ]
    {
    \setlength{\fboxrule}{0.01pt}%
    \scalebox{0.6}{$\kappa=#2$}
    };
    }

\newcommand{\gPlot}[2]{
\begin{gathered}\includegraphics[width=#1\textwidth]{figures/efp_#2.pdf}\end{gathered}
}
\begin{document}

\def\UrlBreaks{\do\/\do-}

\thispagestyle{empty}
\vspace*{-3.5cm}

\vspace{0.5in}

\title{Learning to Isolate Muons}

\author{Julian Collado}
\affiliation{Department of Computer Science, University of California, Irvine, CA, 92697}
\author{Kevin Bauer}
\affiliation{Department of Physics and Astronomy, University of
  California, Irvine, CA 92697}
\author{Edmund Witkowski}
\affiliation{Department of Physics and Astronomy, University of
  California, Irvine, CA 92697}
\author{Taylor Faucett}
\affiliation{Department of Physics and Astronomy, University of
  California, Irvine, CA 92697}
\author{Daniel Whiteson}
\affiliation{Department of Physics and Astronomy, University of
  California, Irvine, CA 92697}
\author{Pierre Baldi}
\affiliation{Department of Computer Science, University of California, Irvine, CA, 92697}

    \begin{abstract}
    Distinguishing between prompt muons produced in heavy boson decay and muons produced in association with heavy-flavor jet production is an important task in analysis of collider physics data.  We explore whether there is information available in calorimeter deposits that is not captured by the standard approach of isolation cones. We find that convolutional networks and particle-flow networks accessing the calorimeter cells surpass the performance of isolation cones, suggesting that the radial energy distribution and the angular structure of the calorimeter deposits  surrounding the muon contain unused discrimination power. We assemble a small set of high-level observables which summarize the calorimeter information and close the performance gap with networks which analyze the calorimeter cells directly.  These observables are theoretically well-defined and can be \rvv{studied with} collider data. 
      \end{abstract}

\date{\today}

\maketitle

\section{ Introduction}

Searches for new physics and precision tests of the Standard Model at hadron colliders have long relied on leptonic decays of heavy bosons, due to the relatively low background rates and excellent momentum resolution compared to hadronic final states. In the case of muons, the primary source of background to prompt muons (those from $W, Z$ or other bosons) is production within a heavy-flavor jet. This non-prompt background is largest at lower values of muon transverse momentum, which has become important in searches for supersymmetry~\cite{Aaboud:2017leg,SCHOFBECK2016631,Khachatryan:2015kxa} as well as low-mass resonances~\cite{Hoenig:2014}.

The current state of the art strategy for distinguishing prompt and non-prompt muons in experimental searches \rev{involves techniques which integrate information from multiple detector systems~\cite{Sirunyan:2017ulk,Pata:2021oez}. Critical to these strategies is the concept of isolation, which is sensitive to the presence of an associated jet that produces many tracks and calorimeter deposits. While the entire detector is worth studying\rvv{~\cite{CMS:2017yfk}}, here we focus on the nature of the information available in the calorimeter. There, the traditional approach is to use a robust and simple method, measuring:}

\[  I_\mu(R_0) = \sum_{i, R<R_0} \frac{p_{\textrm{T}}^{\textrm{cell } i}}{p_{\textrm{T}}^{\textrm{muon}}} \]

\noindent
within a cone $R = \sqrt{\Delta \phi^2 + \Delta \eta^2} < R_0$ surrounding the muon~\cite{Aad:2016jkr}.\rev{ Typically a single cone is used, with  values of $R_0$ in the 0.1-0.45 range}. This approach \rev{relies} on identifying a typical characteristic of the signal, low calorimeter activity in the vicinity of the muon. 

\rev{The}  traditional strategy, however, focuses on the simple nature of the signal  and may overlook the rich set of characteristics offered by the background object, which can provide handles for  additional rejection power.  Related work, which approaches similar object classification tasks as a background jet rejection problem, has shown significant improvement in background discrimination when applied to photons~\cite{Aaij:2017rft,Hall:2018jub}, pions~\cite{ATL-PHYS-PUB-2020-018} or electrons~\cite{Collado:2020fwm}. Other studies have shown that muons which fail the traditional isolation requirement can contain  power to reveal new physics~\cite{Brust:2014gia}.

At the same time, there have been significant advances in machine learning techniques and their applications in physics~\cite{baldi2014searching,baldi2021deep}, specifically in the context of jet classification tasks, which take a fuller view of the object by directly analyzing the low-level calorimeter energy deposits, representing them either as a type of image~\cite{Cogan:2014oua,Baldi:2016fql} or as a list~\cite{Komiske:2018cqr}.

It seems likely, therefore, that these machine learning strategies  may identify the presence of significant additional calorimetric rejection power  in the context of prompt muon identification. In this paper, we apply machine learning tools similar to those developed for jet calorimeter analysis to the task of distinguishing muons due to heavy boson decay from those produced within a heavy-flavor jet, analyze the nature of the information being used, and assemble a set of interpretable calorimeter features which capture that additional classification power.  

\section{Approach and Dataset}

The  observable $I_\mu(R_0)$ is a powerful discriminator which reduces a large amount of information to a single high-level scalar. However, it is possible that it fails to capture the fullness of the  calorimeter information available to distinguish prompt muons from those which are produced within a jet.  To probe whether information has been lost, we  compare the performance of deep neural networks which access the full calorimeter information to shallow networks which use one or more isolation cones.

Neural network decisions are notoriously difficult to reverse-engineer\rvv{~\cite{Chang:2017kvc,Baldi:2014kfa,Roxlo:2018adx,Wunsch:2018oxb,Agarwal:2020fpt}, especially when the dimensionality of the data is large, as is the case for networks which directly use the low-level calorimeter cells. Understanding the nature of the decisions is particularly vital when the training is done with simulated samples, as it leads to valid concerns about the application of such complex strategies to collider data.}

In this study, our goal is not to develop deep networks  for use in collider data. Instead, we apply these deep networks as a probe, to measure a loose upper bound on the possible classification performance, and provide insight into whether information has been lost in the reduction of the calorimeter cells to isolation cones.

Where information has been lost, we attempt to capture it, not by applying the deep network, but by assembling a small set of new high-level (HL) observables that bridge the performance gap and reproduce the classification decisions of the calorimeter cell networks~\cite{faucett}. These high-level observables are more compact, physically interpretable, can be validated in data, and allow \rvv{for} the straightforward assessment and propagation of  systematic uncertainties.

\subsection{Data generation} 

Samples of simulated prompt muons were generated via the process $p p \rightarrow Z'\rightarrow \mu^+ \mu^-$ with a $Z'$ mass of $20$~GeV. Non-prompt muons were generated via the process $p p \rightarrow b \bar{b}$. Both samples are generated at a center of mass energy $\sqrt{s}= 13$~TeV. Collisions and heavy boson decays are simulated with {\sc Madgraph5} \rvv{v2.6.5} ~\cite{madgraph}, showered and hadronized with {\sc Pythia} \rvv{v8.235}~\cite{pythia}, and the detector response  simulated with {\sc Delphes} v3.4.1~\cite{delphes} using the standard ATLAS card \rvv{and {\sc root} version 6.08\/00 \cite{ROOT}}. The classification of these objects is sensitive to the presence of additional proton interactions, referred to as pile-up events. We overlay such interactions within the simulation with an average number of interactions per event of $\mu = 50$, as an estimate of  \rvv{LHC Run 2} experimental data. 

\begin{figure}
\centering
\includegraphics[width=0.75\linewidth]{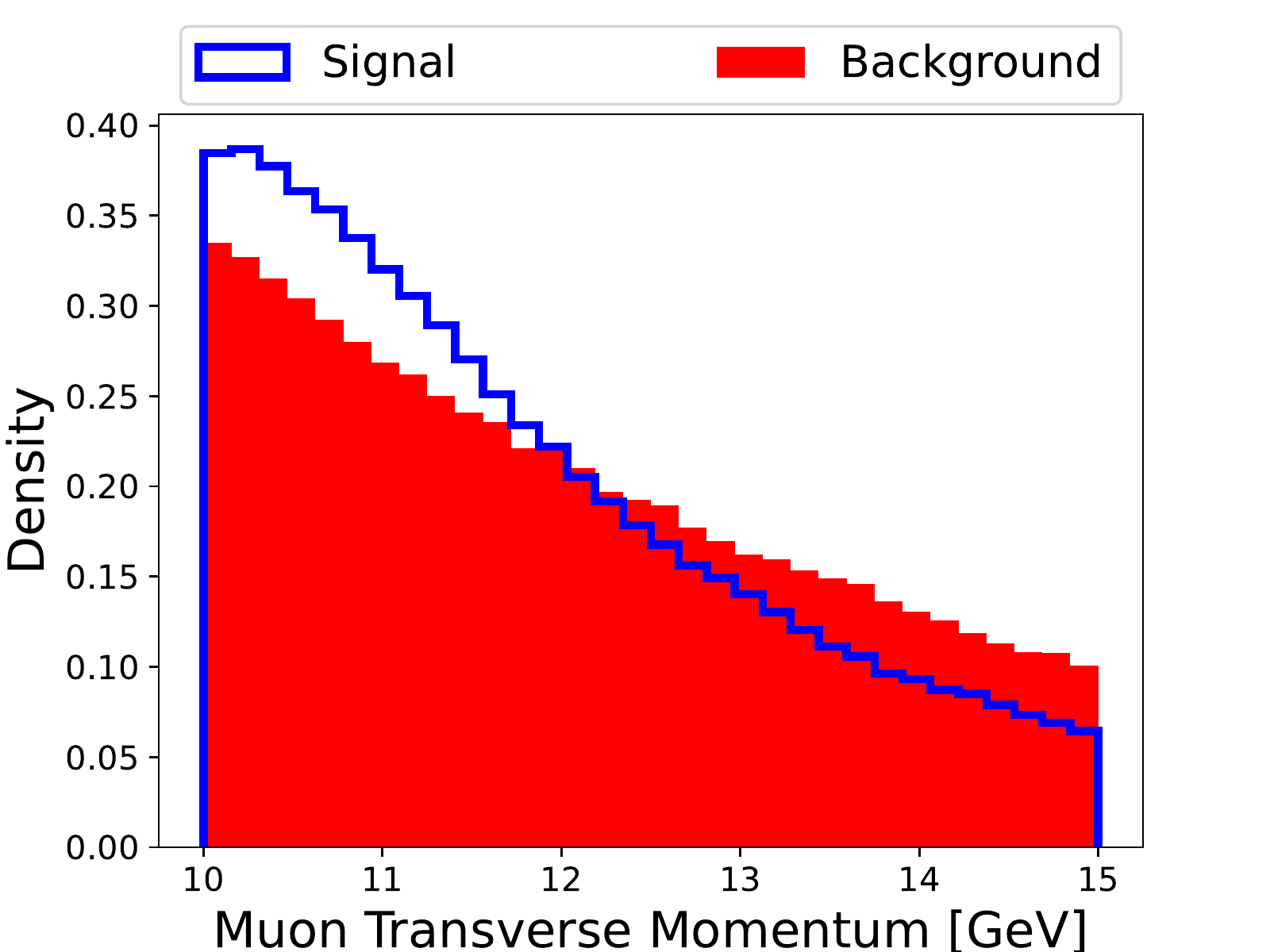}%
\label{fig:pT_dists} \\
(a) \\
\includegraphics[width=0.75\linewidth]{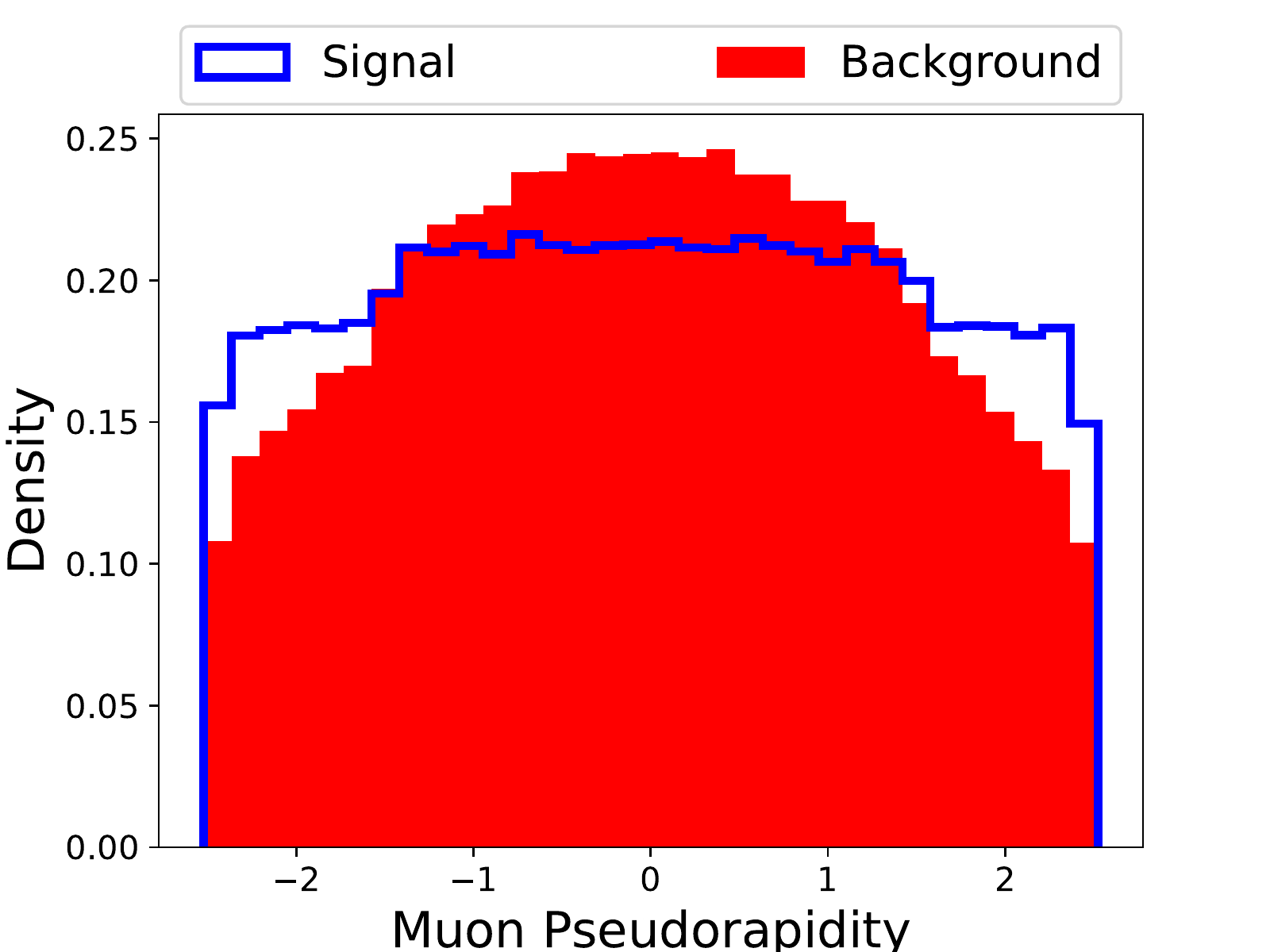}%
\label{fig:eta_dists} \\
(b) \\
\caption{ \rev{ Distributions of muon transverse momentum (top) and pseudorapidity (bottom) for signal and background samples. 
Afterwards, the distributions are weighted to make both samples uniform.}}
\label{fig:weighted_dists}
\end{figure}

Muons in the range $p_{\textrm{T}} \in [10,15]$~GeV \rev{with $|\eta|<2.53$ were considered; \rvv{see Fig.~\ref{fig:weighted_dists}}. To avoid inducing biases from artifacts of the generation process, signal and background events are weighted such that the distributions in $p_\textrm{T}$ and $\eta$ are uniform, using 32 bins in each dimension.} Only events where a muon is identified as a track in the muon spectrometer are used. In total, \rev{499,970 events were used, where 249,991 were signal and 249,979 were background. Both the signal and background datasets are randomly split as: 83\% training, 8.5\% validation, and 8.5\% testing sets.} 

Calorimeter deposits can be represented as images where each pixel value represents the $E_\textrm{T}$ deposited by a particle~\cite{Cogan:2014oua}.  Images are formed by considering cells in the calorimeter within a cone of radius up to $\Delta R = 0.45$ surrounding the muon location after \rev{propagating} to the radius of the calorimeter.

\begin{figure}
\centering
\includegraphics[width=0.75\linewidth]{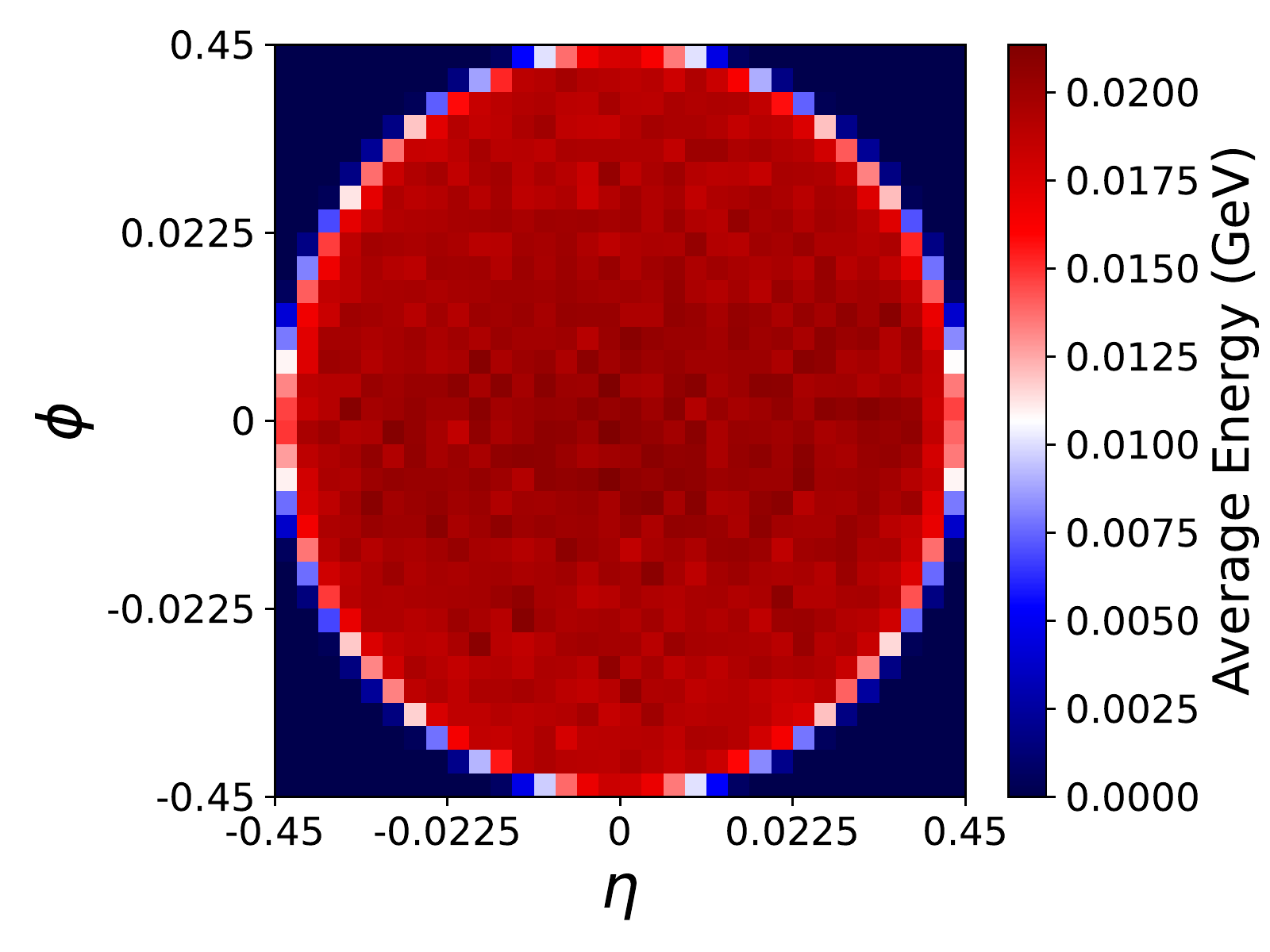}%
\label{fig:sig_avg} \\
(a) Mean Prompt Muon\\
\includegraphics[width=0.75\linewidth]{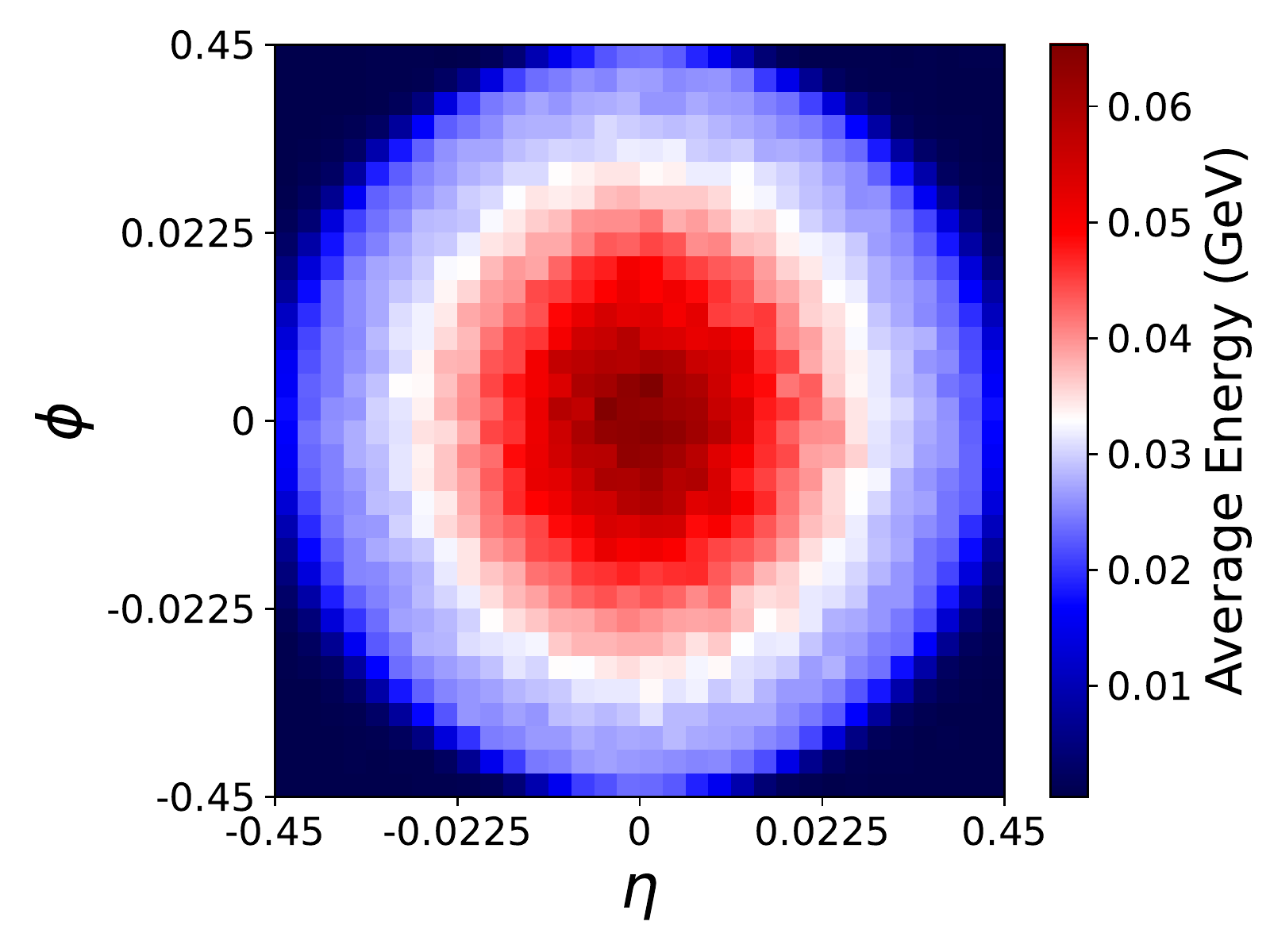}%
\label{fig:bg_avg} \\
(b) Mean Non-prompt Muon \\
\caption{ Mean calorimeter images for signal prompt muons (top) and muons produced within heavy-flavor jets (bottom), in the vicinity of reconstructed muons within a cone of $R=0.45$. The color of each cell represents the sum of the $E_\textrm{T}$ of the calorimeter deposits within the cell. }
\label{fig:dep}
\end{figure}

We use a 32x32 grid, which approximately corresponds with the calorimeter granularity of ATLAS and CMS. Heat maps of the calorimeter energy deposits in $\eta-\phi$ space for both signal prompt muons and background non-prompt muons are shown in Fig.~\ref{fig:dep}. The signal calorimeter deposits are uniform and can be attributed to pileup whereas the background deposits appear largely radially symmetric with a dense core from the jet.

We calculate the standard muon isolation observable $I_\mu(R_0)$ for a set of cones with $0.025 \leq R_0 \leq 0.45$ in 18 equally spaced steps. 

Crucially, these isolation observables and all other calorimeter observables are calculated directly from the pixels of the muon images, ensuring that they contain a strict subset of the information available. This allows for direct and revealing comparisons of the performance between networks trained with the images and those trained with $I_\mu$. Note that pixelization of the detector may incur some loss of information relative to the underlying segmentation of the calorimeter. \rev{ However, this work focuses on examining the relative power of different techniques, rather than identifying the best performance under the most realistic scenario.}

\section{Networks and Performance}

\begin{figure}
\includegraphics[width=0.95\linewidth]{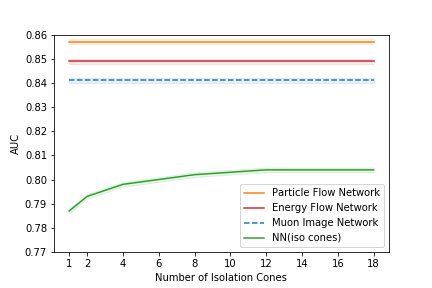}
\caption{ Comparison of classification performance using the  metric  AUC  between Particle-Flow networks trained on  lists of calorimeter deposits (orange, solid), \rvv{Energy-Flow networks trained on lists of calorimeter deposits (red, solid),} convolutional networks trained on muon images (blue, dashed) and networks which use increasing numbers of isolation cones (green, solid). For each number of cones, the optimal set is chosen.
}
\label{fig:iso_comp}
\end{figure}

\begin{figure}
\includegraphics[width=0.95\linewidth]{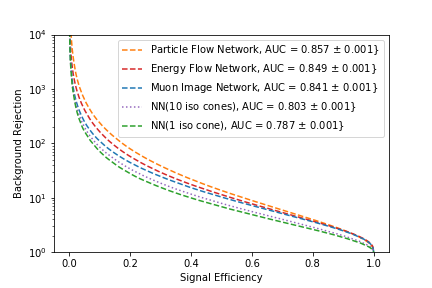}
\caption {Background rejection versus signal efficiency for Particle-Flow networks trained on  lists of calorimeter deposits (orange, \rvv{dashed}), \rvv{Energy-Flow networks trained on lists of calorimeter deposits (red, dashed),} convolutional networks trained on muon images (blue, dashed), networks trained on a set of isolation cones (purple, dotted) and the benchmark approach, a single isolation cone approach
(green, dashed).
}
\label{fig:iso_roc}
\end{figure}

We apply several strategies to the task of classifying prompt and non-prompt muons, \rvv{using both low-level calorimeter information and higher-level isolation quantities.   We evaluate the performance of each approach by comparing the integral of the ROC (Receiver Operating Characteristic) curve, known as the AUC (Area Under the Curve). \rev{The uncertainty for the AUC is calculated by training 100 randomly initialized models with the same hyperparameters on different bootstraps of the data. In this case, we seek to determine the statistical uncertainty due to the stochastic training method, rather than any systematic uncertainty due to the calorimeter resolution.} 

For the high-level quantities, the standard approach of using a single isolation cone yields an AUC of \rev{\ISOOneresult} for the optimal cone size, $R_0 = 0.425$\footnote{Similar performance was seen for other cone sizes.}. We hypothesized that additional cones would provide useful information about the radial energy distribution. Including a second cone with a distinct $R_0$ value as input to a small neural network (see Appendix A) slightly improves performance, with an AUC of \rev{\ISOTworesult}. To estimate the full information available in the cones, we perform a greedy search through all 18 cones; we find that a set of {10} cones\footnote{ {$R_0 = [0.025, 0.05,  0.075,  0.125, 0.15,  0.225, 0.275,  0.325,0.425, 0.45]$}} yields another small boost in classification power up to an AUC of {\ISOTenresult}, as shown in Fig.~\ref{fig:iso_comp}.  Performance was fairly insensitive to the specific choices of cone sizes, and does not grow significantly beyond \rev{10} cones. Feed-forward dense networks are trained to use the information in one or more isolation cones (see the Appendix for details on network architectures and training).  

We next examine whether additional information is available by applying strategies which access  the calorimeter information at the lowest-level and highest-dimensionality.  Convolutional networks (CNN) are applied to the muon images~\cite{Cogan:2014oua,Baldi:2016fql,baldi2021deep}. As an alternative, we apply particle-flow networks (PFN)~\cite{Komiske:2018cqr}, which are mathematically structured as sums over inputs and thus are invariant to permutations of the inputs. 

 The muon image CNN achieves a significantly higher performance than the isolation-only networks, with an AUC of \rev{\CNNresult}, and the particle flow network reaches \rev{\PFNresult}, see Fig.~\ref{fig:iso_roc} and Table~\ref{tab:summary}. This  immediately suggests that there is significant additional information available to distinguish between the prompt and non-prompt muons beyond what is summarized in the isolation cones.  A more restricted version of the PFN, an Energy-Flow Network~\cite{Komiske:2018cqr} (EFN), which enforces infra-red and collinear (IRC) safety, achieves nearly the same performance, \rev{0.849}.  }
This suggests that most of the additional information beyond the isolation cones is IRC-safe.

These results support the conventional wisdom that a significant fraction of the information relevant for classification is captured by a single, simple cone. However, they also indicate that there is additional information in the radial distribution of energy, which can be captured by using multiple cones.  \rev{However,} even many cones fail to match the performance of the networks which use the calorimeter cell information directly, suggesting that there is additional non-radial information relevant to the classification task not captured in the isolation cones. \rev{This is likely due to a difference between the muon axis, the center of the isolation cones, and the jet axis.} 

\section{Analysis}
\label{section:MuonImageAnalysis}

The networks which use the calorimeter cells directly have the most powerful performance, but our aim is not simply to optimize  classification performance in this particular simulated sample. Instead, we seek to understand the nature of the learned strategy in order to validate it and translate it into simpler, more easily interpretable high-level features which can be studied in other datasets, real or simulated.  In addition, this understanding can reveal how well the strategy is likely to generalize to other kinds of jets that are not represented by this background sample, such as charm jets.

The CNN and PFN results indicate that the radially symmetric isolation cones are failing to utilize some  information which is relevant to the classification task.  In this section, we search for additional high-level observables which  capture this information. 

\subsection{Search Strategy}

Interpreting the decisions of a deep network with a high-dimensional input vector is notoriously difficult.  Instead, we attempt to translate its performance into a smaller set of interpretable observables~\cite{faucett}. This allows us to understand the nature of the information being used as well as to represent it more compactly.

\rvv{One might imagine exploring a set of physically-motivated quantities, such as the relative $p_\text{T}$ between the jet and the muon or the energy-weighted average distance between the jet and calorimeter cells. These particular quantities were considered and found to not contribute  significant power in addition to the isolation cones. 

Instead, we use a systematic approach and explore a formally complete set of observables}. As the background non-prompt muons are due to jet production,  we search within a set of observables originally intended for analysis of jets: the Energy Flow Polynomials (EFPs)~\cite{Komiske:2017aww}, a formally infinite set of parameterized engineered functions, inspired by previous work on energy correlation functions \cite{Larkoski:2013eh}, which sum over the contents of the cells scaled by relative angular distances. \rev{An EFP  for a jet with $M$ \rvv{constituents} which considers $N$ correlators with angular connections $k,l$ is written as:

\[ \textrm{EFP} = \sum_{i_1=1}^M ... \sum_{i_N=1}^M z^\kappa_{i_1}...z^\kappa_{i_N}\prod_{k,l} \theta^\beta_{i_k i_l} \]

\noindent where

\begin{align}
	(z_i)^\kappa &= \left(\frac{p_{\textrm{T}i}}{\sum_j p_{\textrm{T}j}} \right)^\kappa, \label{eq:EFP_z} \\
	\theta^\beta_{ij} &= \left(\Delta \eta_{ij}^2 + \Delta \phi_{ij}^2 \right)^{\beta/2}. \label{eq:EFP_theta}
\end{align}

Here, $p_{\textrm{T}i}$ is the transverse momentum of cell $i$, and $\Delta\eta_{ij}$ ($\Delta\phi_{ij}$) is the pseudorapidity (azimuth) difference between cells $i$ and $j$.  These parametric sums correspond to the set of all isomorphic multigraphs where:
\begin{align}
	\text{each node} &\Rightarrow \sum_{i = 1}^N z_i, \label{eq:EFP_node}  \\
	\text{each $k$-fold edge} &\Rightarrow \left(\theta_{ij}\right)^k \label{eq:EFP_edge} . 
\end{align}
}
 As the EFPs are normalized, they capture only the relative information about the energy deposition. For this reason, in each network that includes EFP observables, we include as an additional input the sum of $p_{\textrm{T}}$ over all cells, to indicate the overall scale of the energy deposition.

The original IRC-safe EFPs require $\kappa = 1$. To more broadly explore the space,  we consider examples with $\kappa \not= 1$ to explore a broader space of observables\footnote{Also, note that $\kappa > 0$ generically corresponds to IR-safe but C-unsafe observables. For $\kappa < 0$, empty cells are omitted from the sum.}. 

 In principle, the space spanned by the EFPs is complete, such that any jet observable can be described by one or more EFPs of some degree. One might consider simply searching this space for all possible combinations of EFPs for a set which maximizes performance for this task. Such a search is computationally prohibitive; instead, we follow the black-box guided algorithm of Ref.~\cite{faucett}, which iteratively assembles a set of EFPs that mimic the decisions of another guiding network (the PFN in our case) by isolating the portion of the input space where the guiding network disagrees with the isolation network, and finding EFPs which mimic the guiding network's decisions in that subspace.
 
 Here, the agreement between networks $f(x)$ and $g(x)$ is evaluated over pairs of $(x,x')$ by comparing their relative classification decisions, expressed mathematically as: 
 
\begin{equation}
\label{eq:DO_def}
	\textrm{DO}[f,g](x,x') = \Theta \Big( \big( f(x) - f(x') \big) \big( g(x) - g(x') \big) \Big),
\end{equation}

\noindent
and referred to as {\it decision ordering} (DO). A DO$=0$ corresponds to inverted decisions over all input pairs and DO$=1$ corresponds to the same decision ordering.  As prescribed in Ref.~\cite{faucett}, we scan the space of EFPs to find the observable that has the highest average decision ordering (ADO) with the guiding network when averaged over  disordered pairs.  The selected EFP is then incorporated into the new network of HL features, HLN$_{n+1}$, and the process is repeated until the ADO  plateaus.

\subsection{IRC Safe Observables}

\rvv{As the elements of the EFP space are not orthogonal, there are potentially many combinations of EFP observables which capture the relevant information. As simpler EFPs may be more conducive to theoretical interpretation, we begin our search in a restricted subset of the EFP space. Specifically, we consider those which are IRC safe ($\kappa=1$), have a simple angular weighting ($\beta \in [1,2]$), and $n\le 3$ fewer nodes  with at most three edges between nodes.}  We also include $\sum p_{\textrm{T}}$, where the summation is over all calorimeter cells in the image, to set the scale accompanying the normalized EFPs. The first EFP observable  identified is a simple three-point correlator: 

\[ \gPlot{0.05}{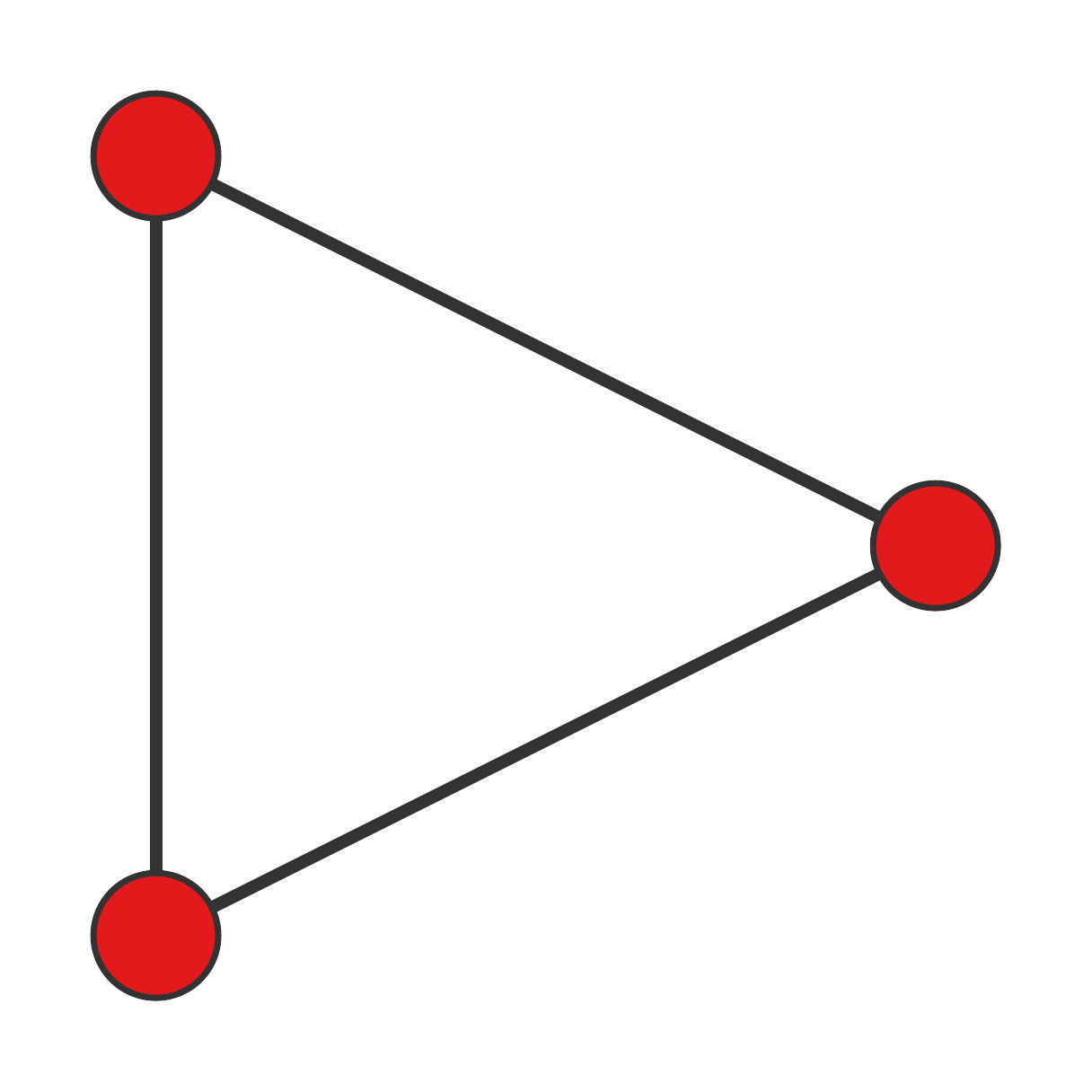} {\scriptstyle (\kappa=1,\beta=1)}= \sum^N_{a,b,c=1}  z_a z_b z_c  \theta_{ab} \theta_{bc} \theta_{ca}  \]

\noindent
which, when combined with the isolation cones and  $\sum p_{\textrm{T}}$, yields an AUC of \rev{0.838} and an ADO with the CNN of \rev{0.891}, a significant boost relative to just using the radial information of the isolation cones. The subsequent scans produce variants of this observable :

\rev{
\[ \gPlot{0.05}{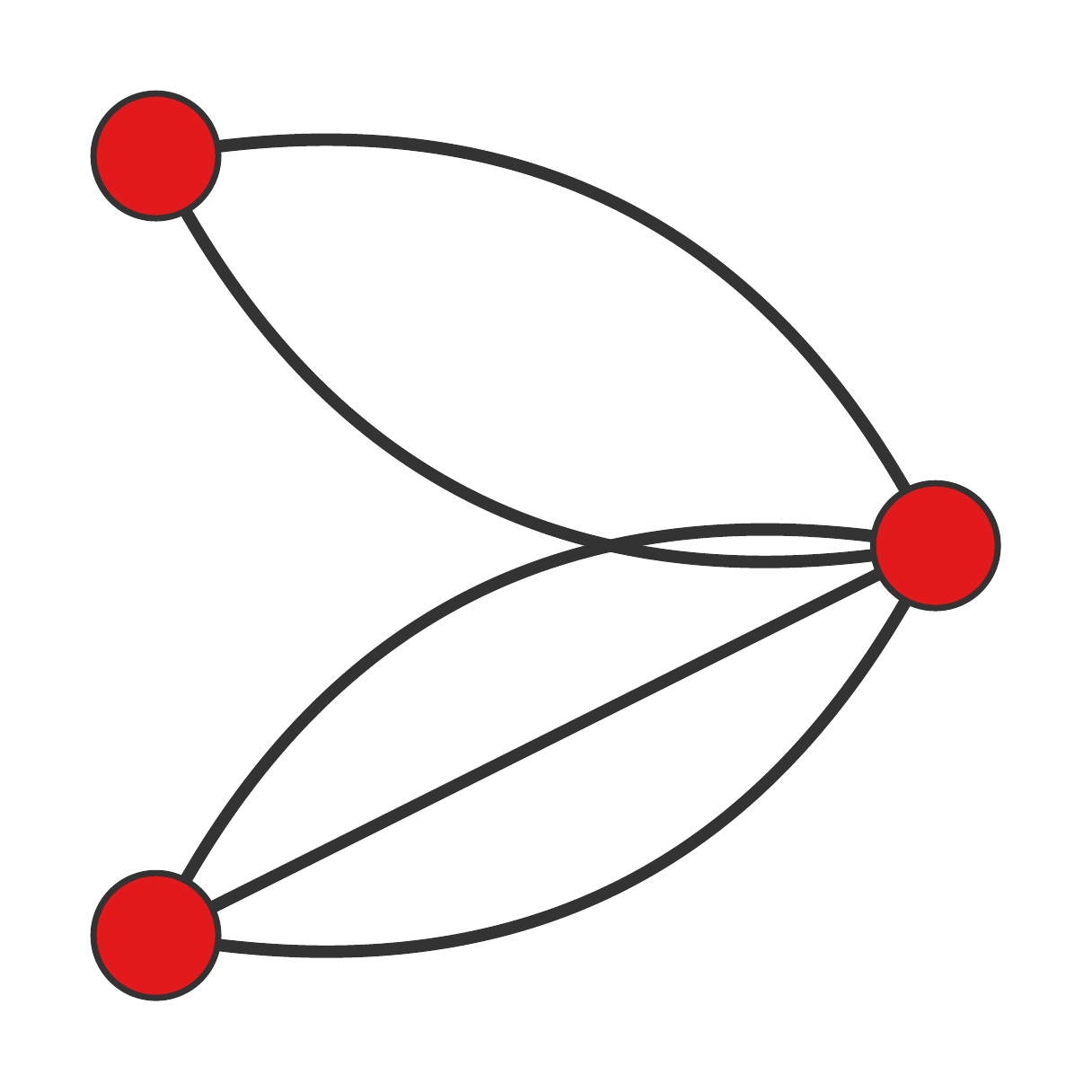} {\scriptstyle (\kappa=1,\beta=2)} = \sum^N_{a,b,c=1}  z_a z_b z_c  \theta_{ab}^4 \theta_{bc}^6  \]

\[ \gPlot{0.05}{3_5_1} {\scriptstyle (\kappa=1,\beta=1)}= \sum^N_{a,b,c=1}  z_a z_b z_c  \theta_{ab}^2 \theta_{bc}^3 \]

\[ \gPlot{0.05}{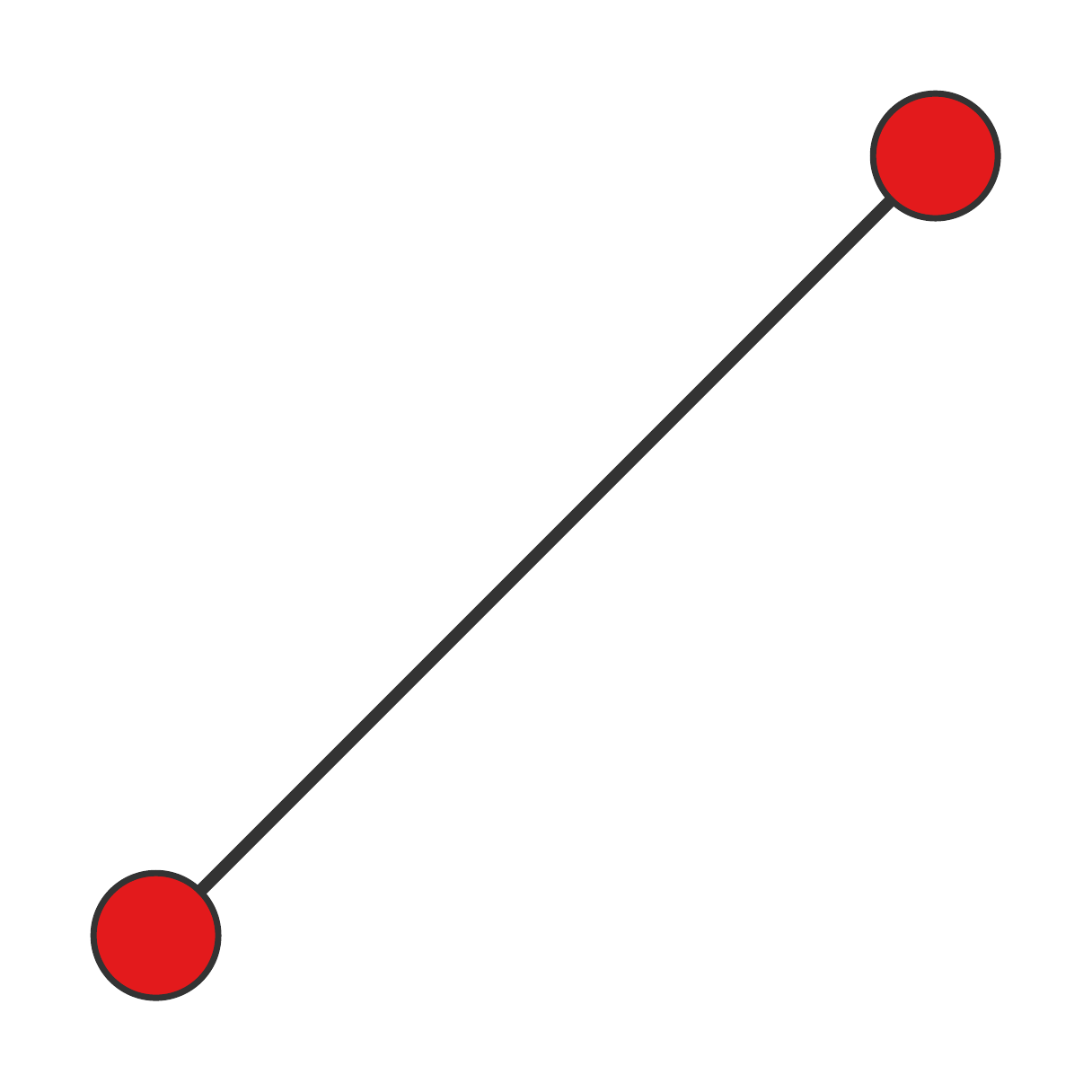} {\scriptstyle (\kappa=1,\beta=2)} = \sum^N_{a,b=1}  z_a z_b   \theta_{ab}\]
}

\noindent 
with additional edges corresponding to higher powers of the angular information. Their power may come from their sensitivity to the collimated radiation pattern of the jet. Together with the isolation cones, these observables reach an AUC of \rev{0.842} and an ADO with the PFN of \rev{0.888}, see Table~\ref{tab:summary}. 

\begin{figure}[h!]
	\centering
    \begin{tikzpicture}
    \node(a){
	\includegraphics[width=0.3\textwidth]{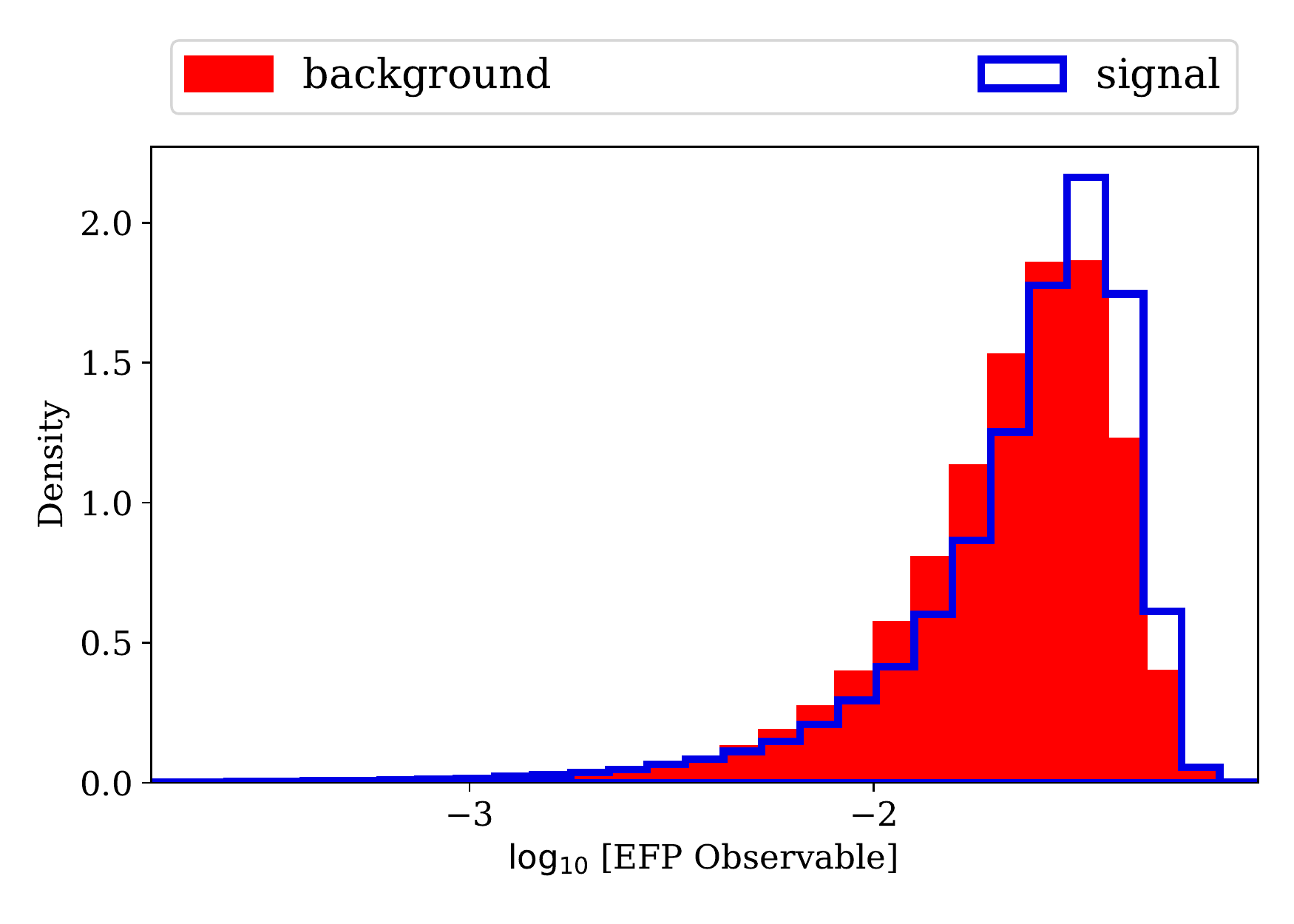}};
    \showefp{3_3_1}{1}{1}
    \end{tikzpicture}
     \begin{tikzpicture}
    \node(a){
	\includegraphics[width=0.3\textwidth]{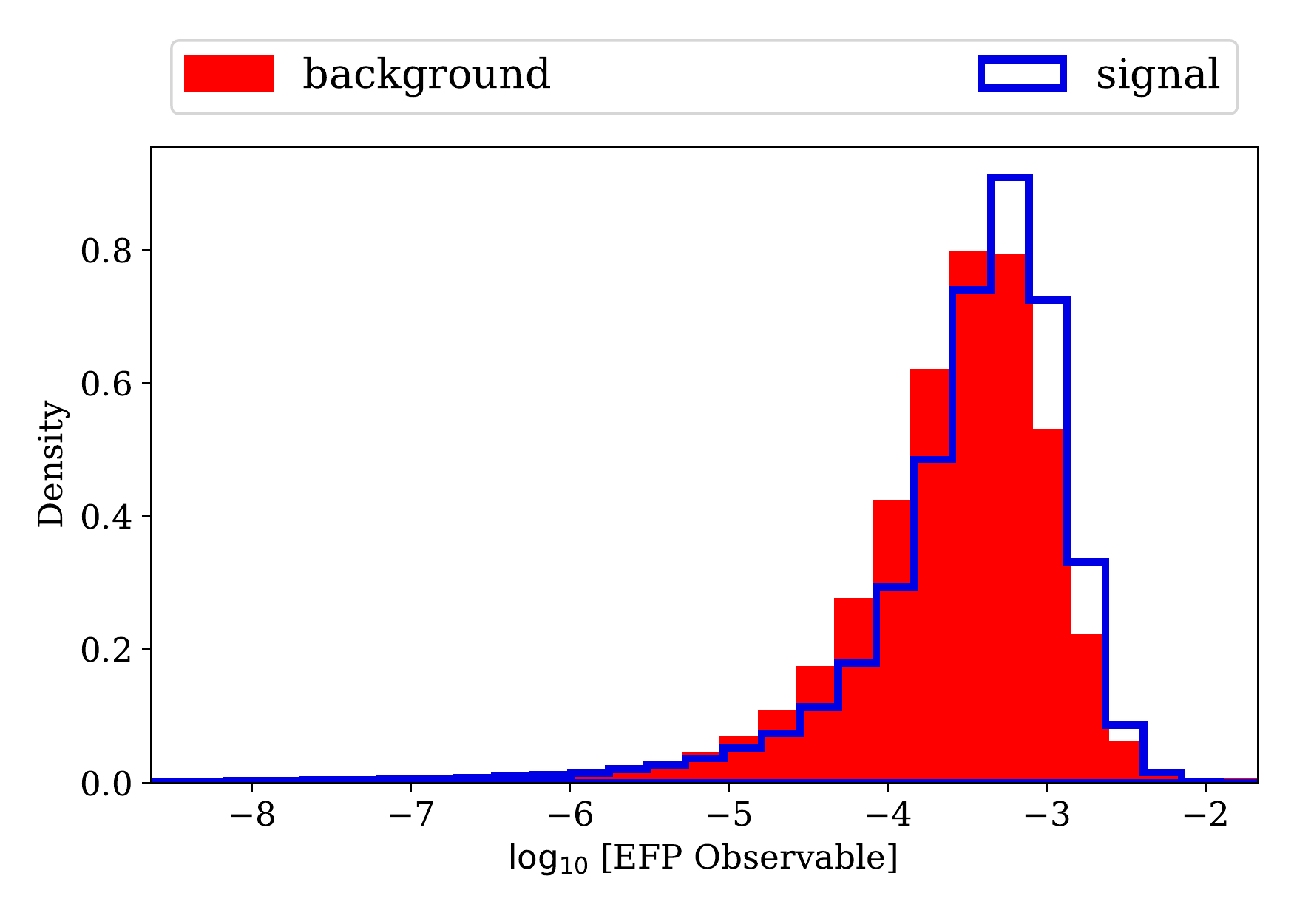}};
	\showefp{3_5_1}{1}{2}
    \end{tikzpicture}
      \begin{tikzpicture}
    \node(a){
	\includegraphics[width=0.3\textwidth]{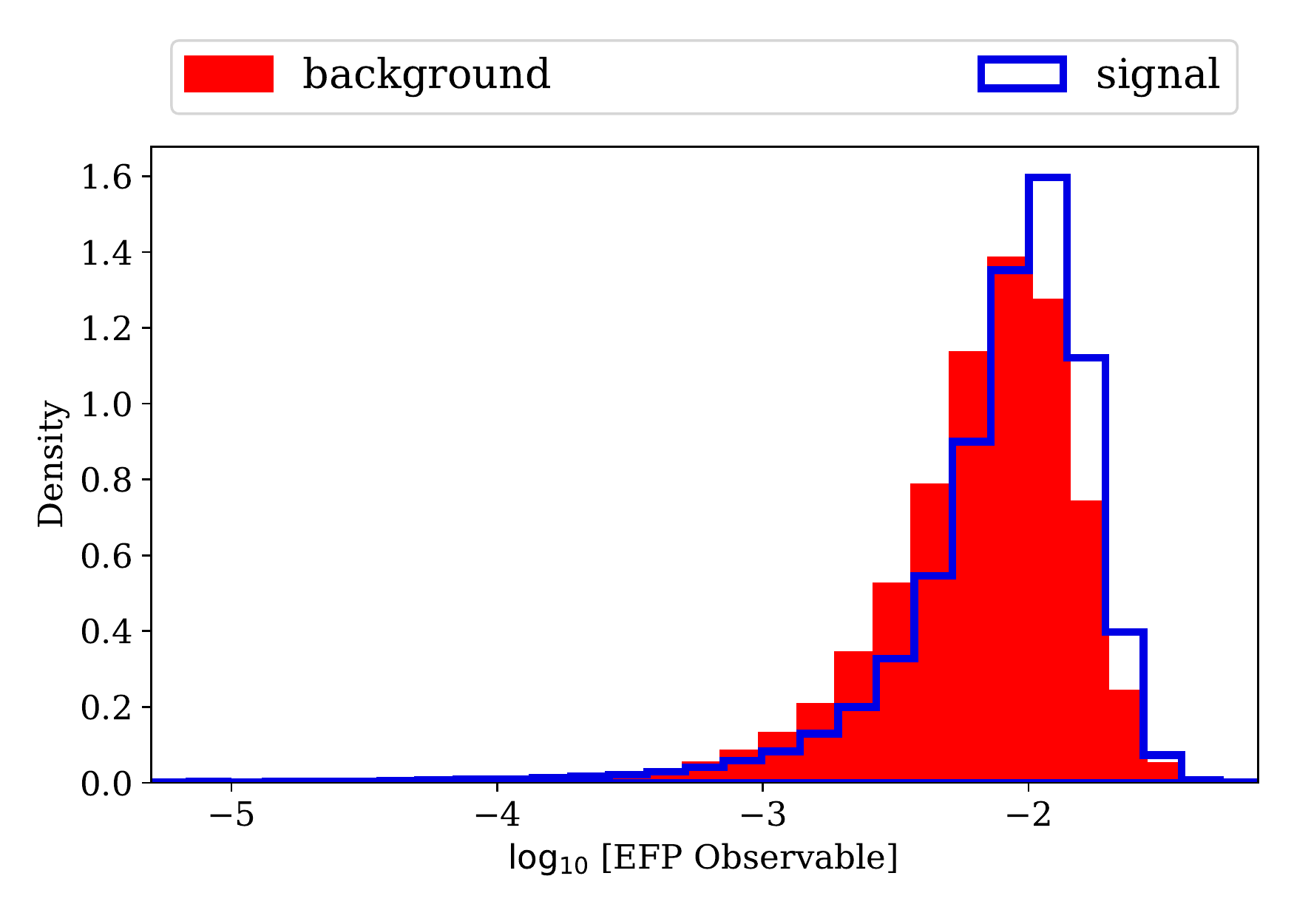}};
    \showefp{3_5_1}{1}{1}
    \end{tikzpicture}
       \begin{tikzpicture}
    \node(a){
	\includegraphics[width=0.3\textwidth]{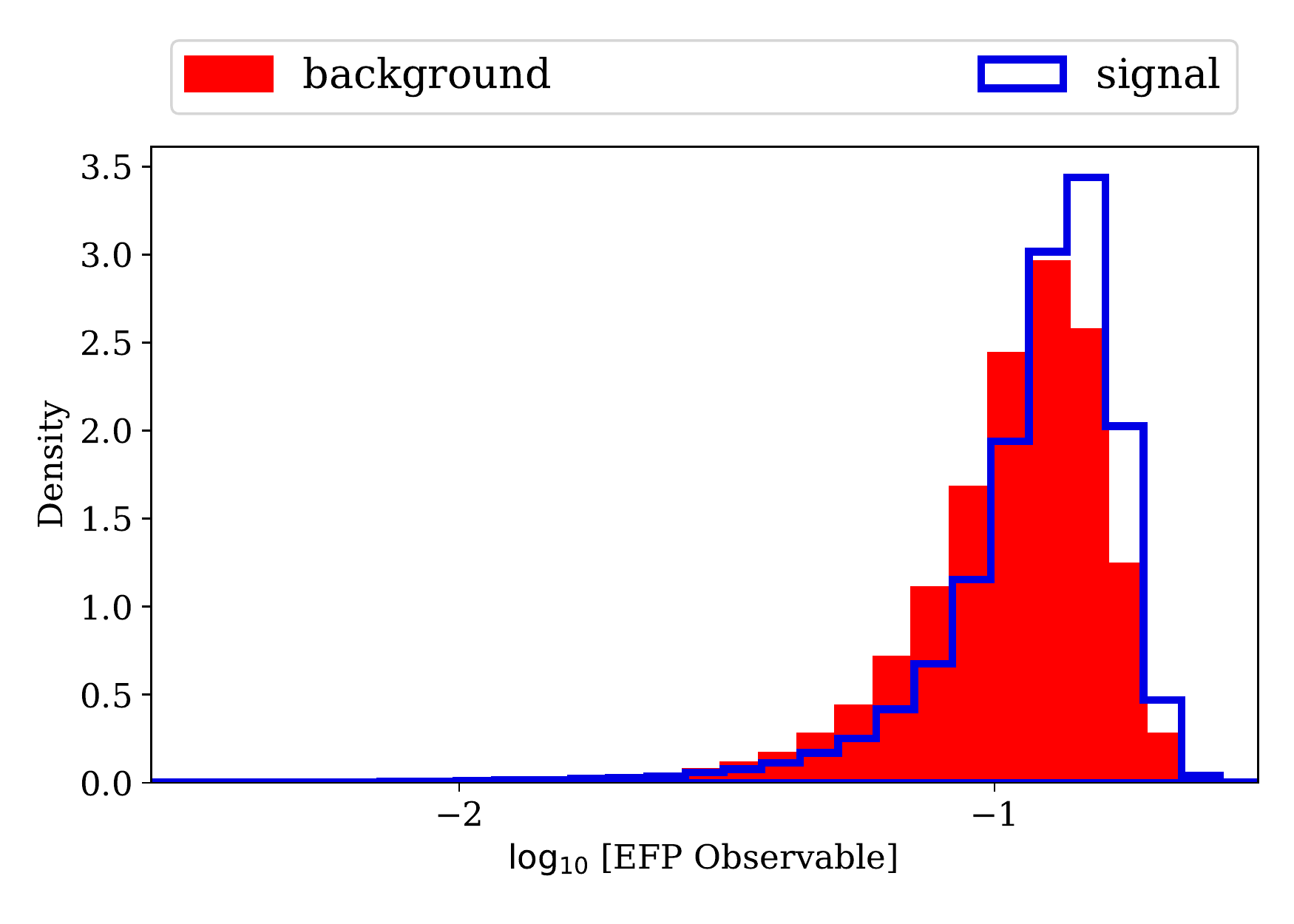}};
    \showefp{2_1_0}{1}{2}
    \end{tikzpicture}
	\caption{Distributions of the $\log_{10}$ of the selected IRC-safe EFPs as chosen by the black-box guided strategy, for prompt (signal) muons and non-prompt (background) muons.}
	\label{fig:efp_safe}
\end{figure}

This set of observables partially closes the performance gap with the best calorimeter cell networks, indicating that angular information is relevant to the muon isolation classification task, but fails to fully match its performance. Distributions of these EFPs for signal and background are shown in Fig.~\ref{fig:efp_safe}. Further scans in this limited space do not yield significant boost in AUC or ADO values. \rev{The strong result of the IRC-safe EFN indicates that it is possible to capture nearly all of the classification power using IRC-safe graphs, likely requiring graphs with complexity beyond what we have considered.  }

A scan guided by the CNN rather than the PFN yields very similar results, with identical choices for the first three EFPs.

\begin{figure}[h!]
	\centering
    \begin{tikzpicture}
    \node(a){
	\includegraphics[width=0.3\textwidth]{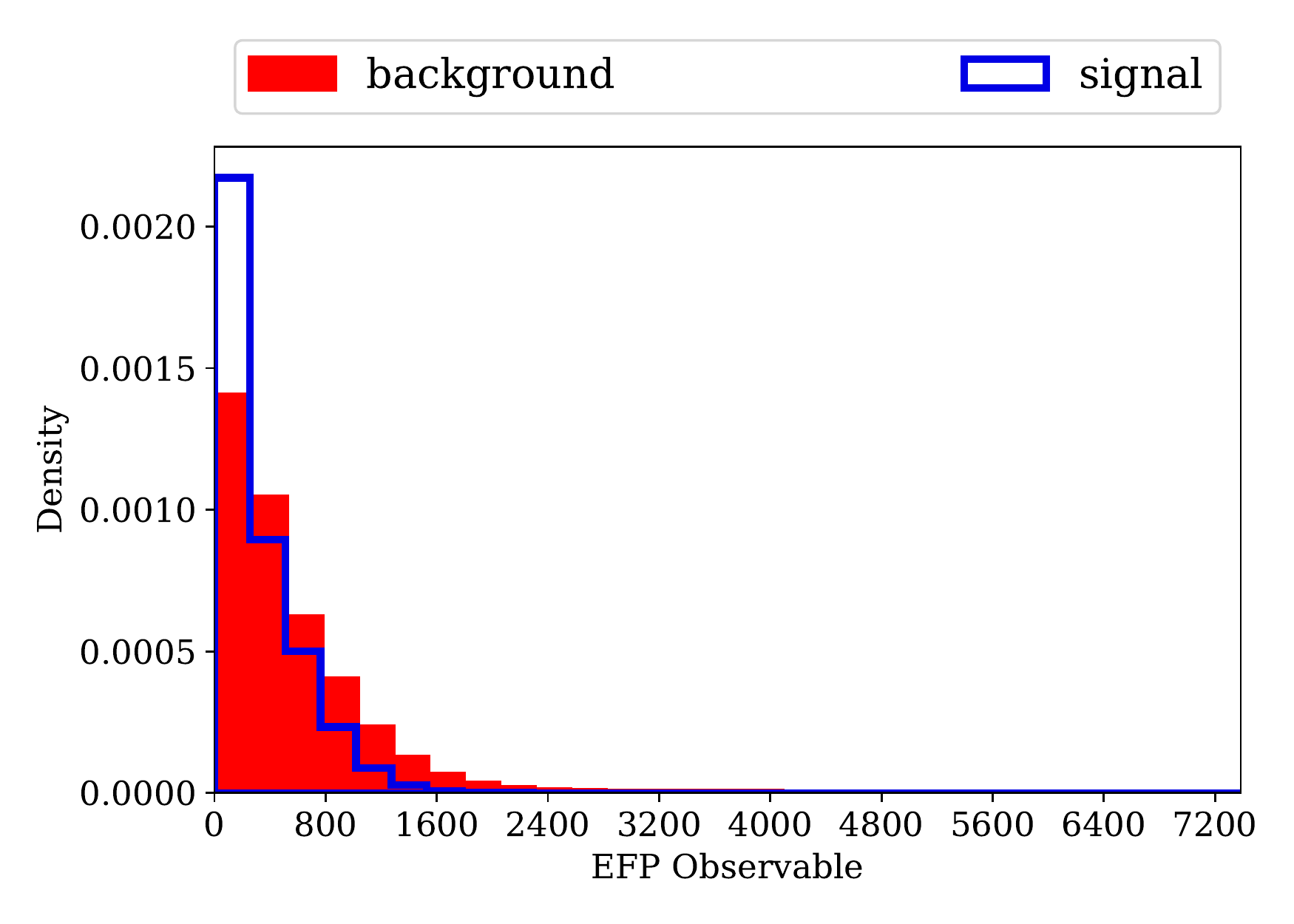}};
    \showefpRk{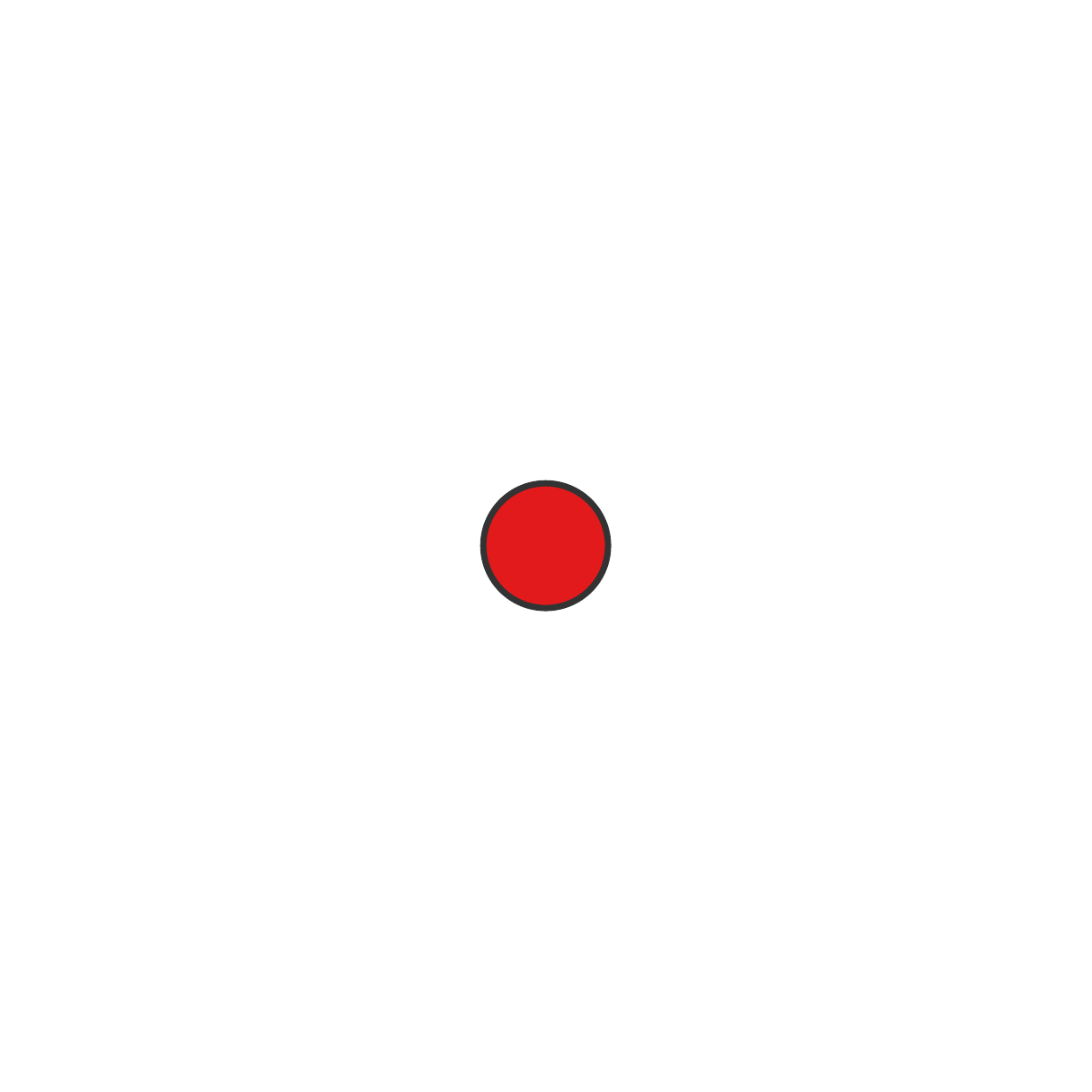}{-1}
    \end{tikzpicture}
     \begin{tikzpicture}
    \node(a){
	\includegraphics[width=0.3\textwidth]{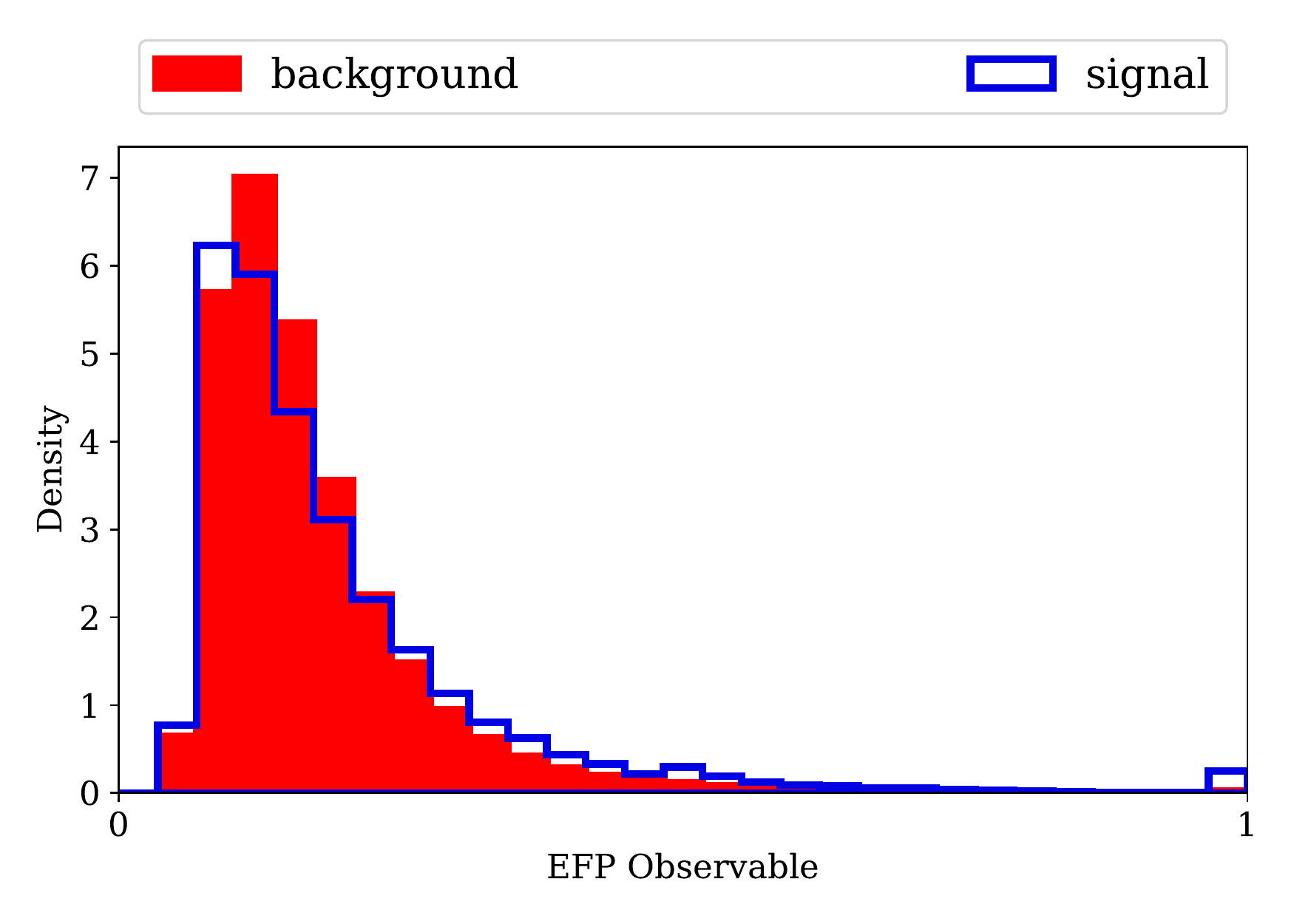}};
	\showefpRk{1_0_0}{2}
    \end{tikzpicture}
      \begin{tikzpicture}
    \node(a){
	\includegraphics[width=0.3\textwidth]{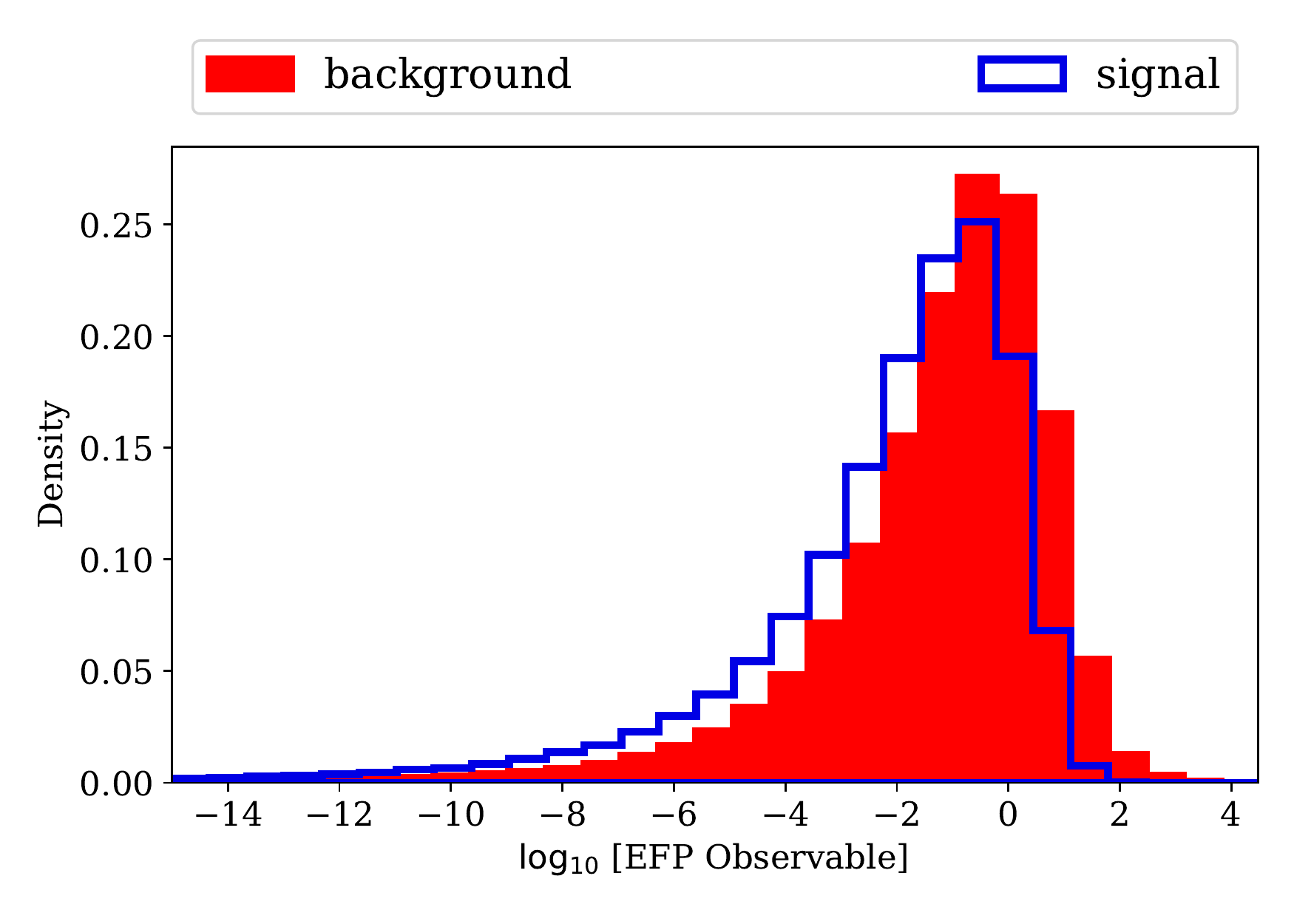}};
    \showefp{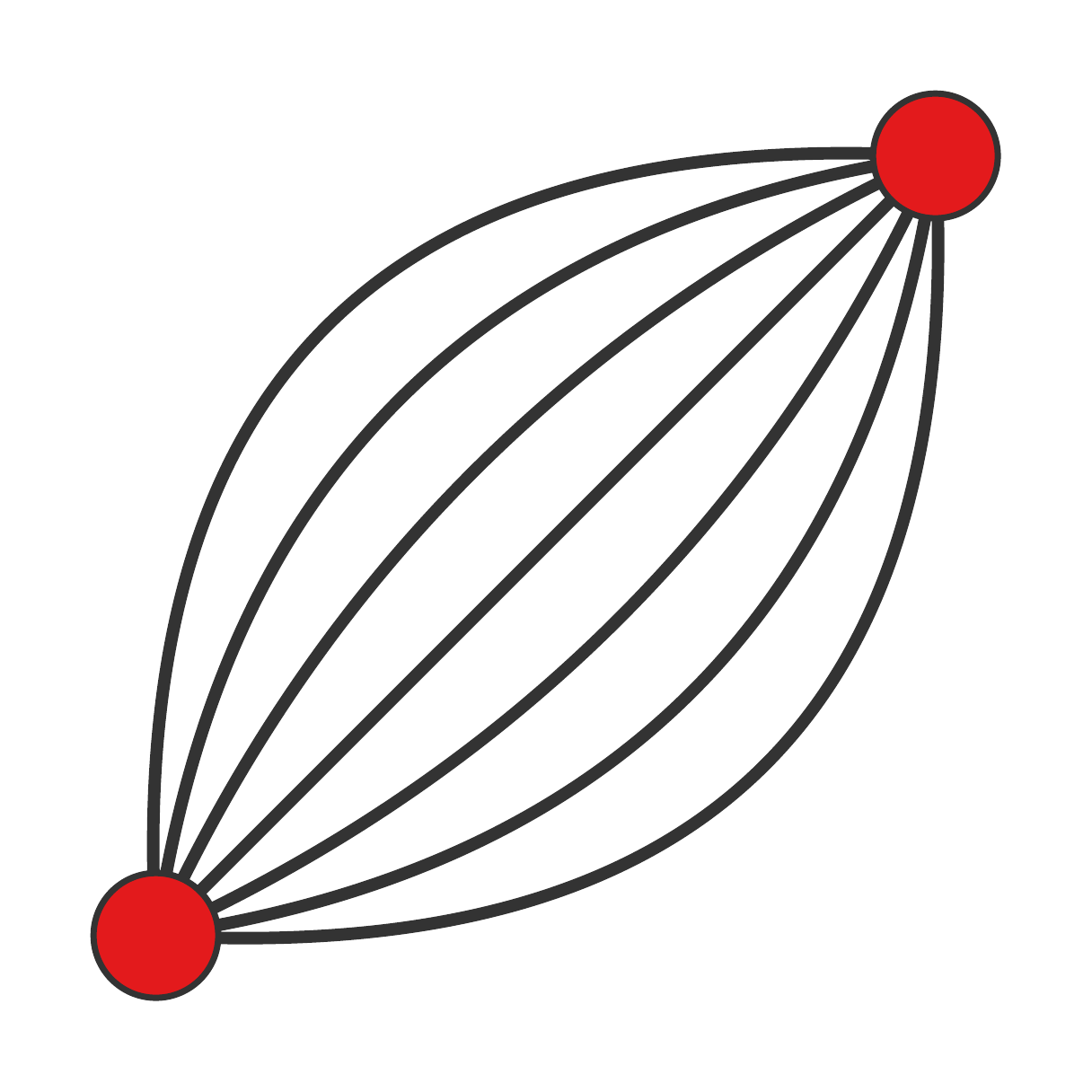}{-1}{4}
    \end{tikzpicture}
       \begin{tikzpicture}
    \node(a){
	\includegraphics[width=0.3\textwidth]{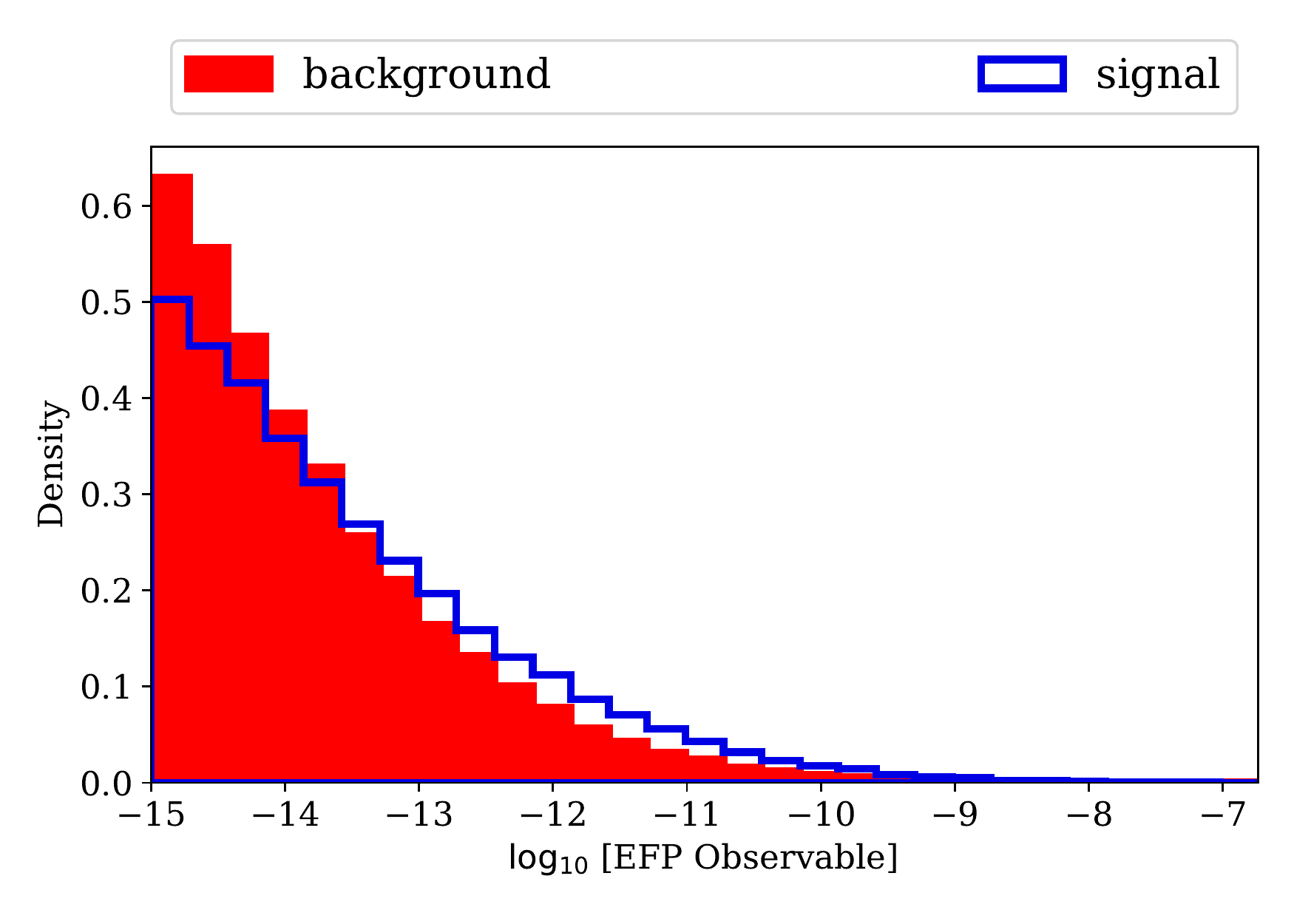}};
    \showefpR{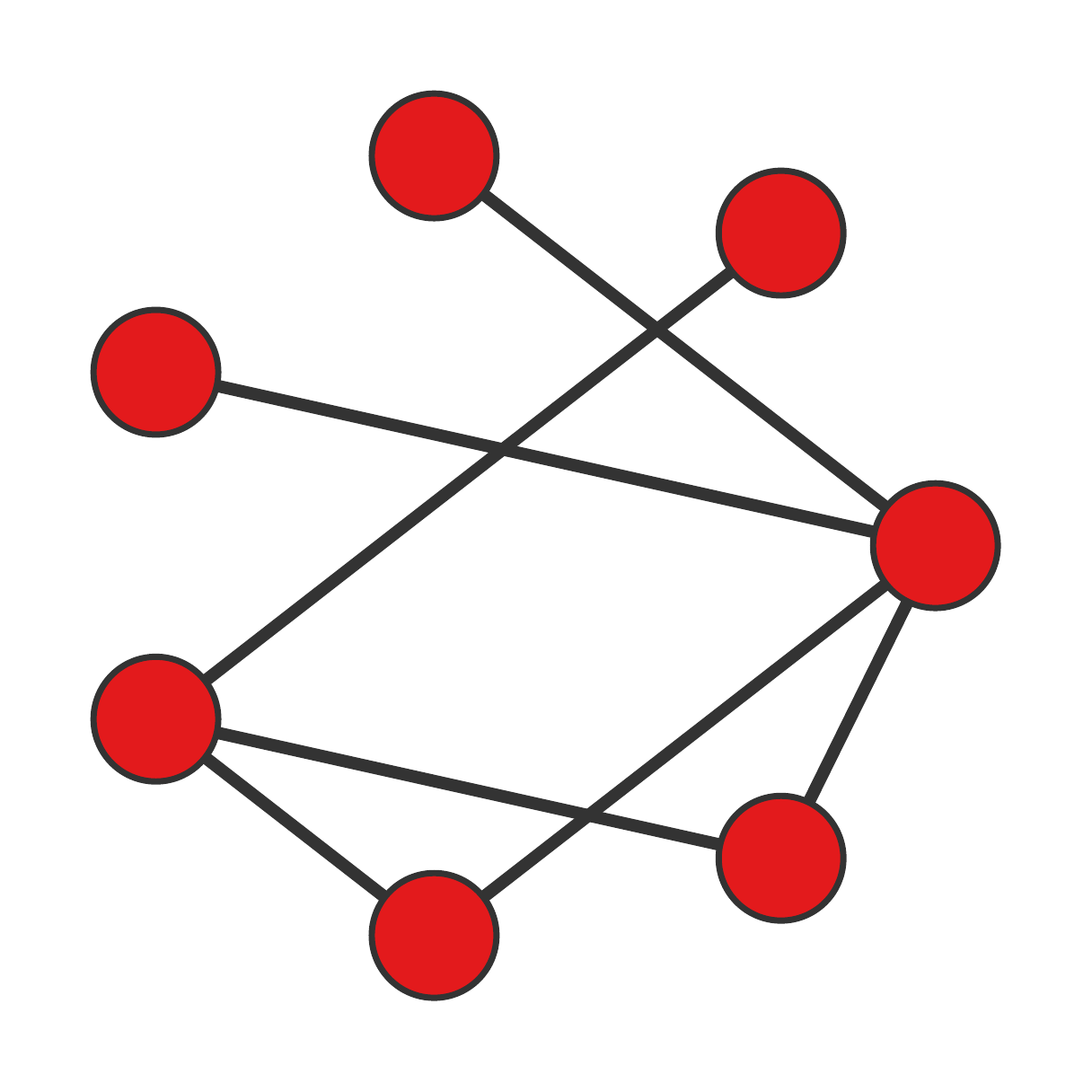}{2}{4}
    \end{tikzpicture}
       \begin{tikzpicture}
    \node(a){
	\includegraphics[width=0.3\textwidth]{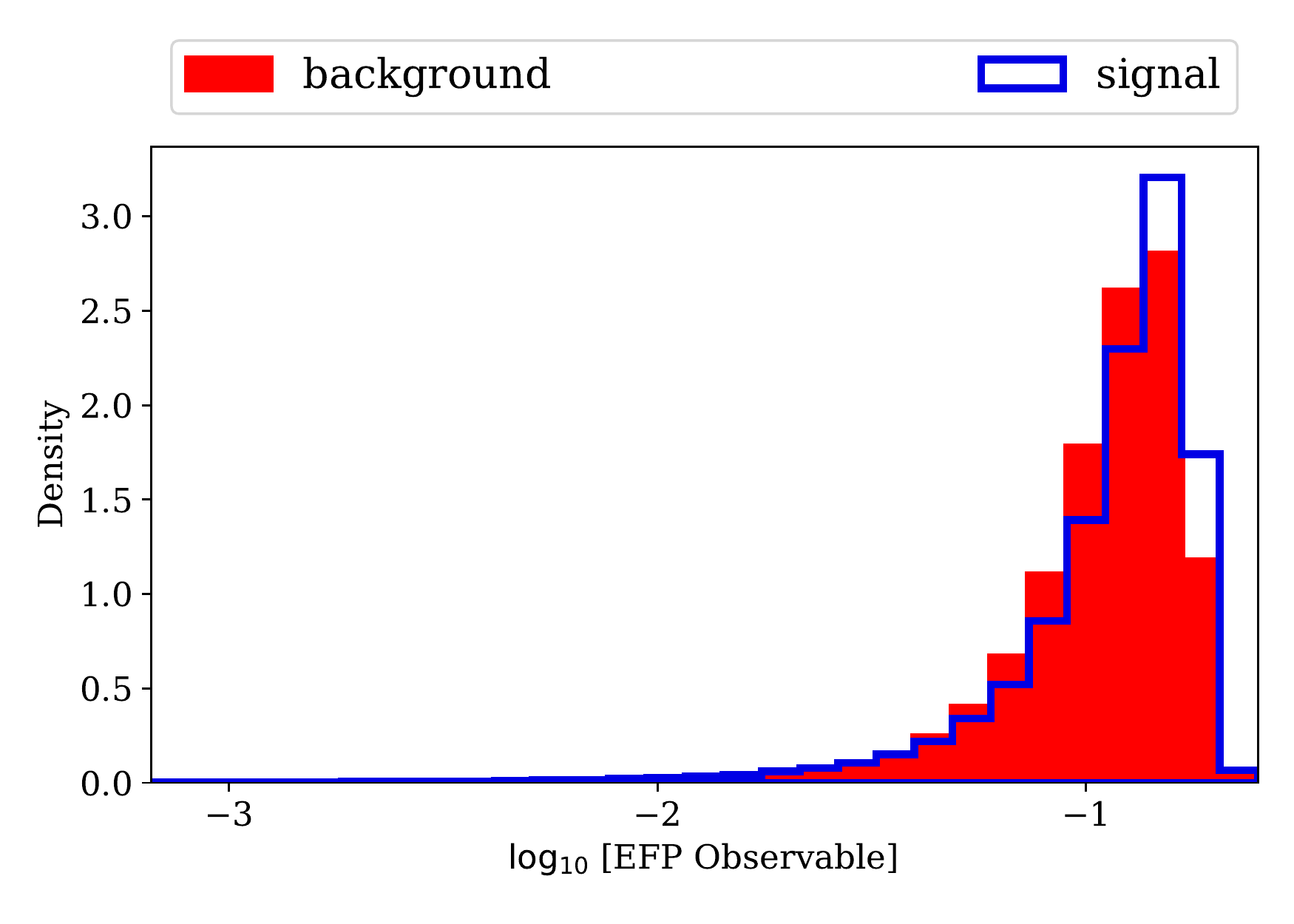}};
    \showefp{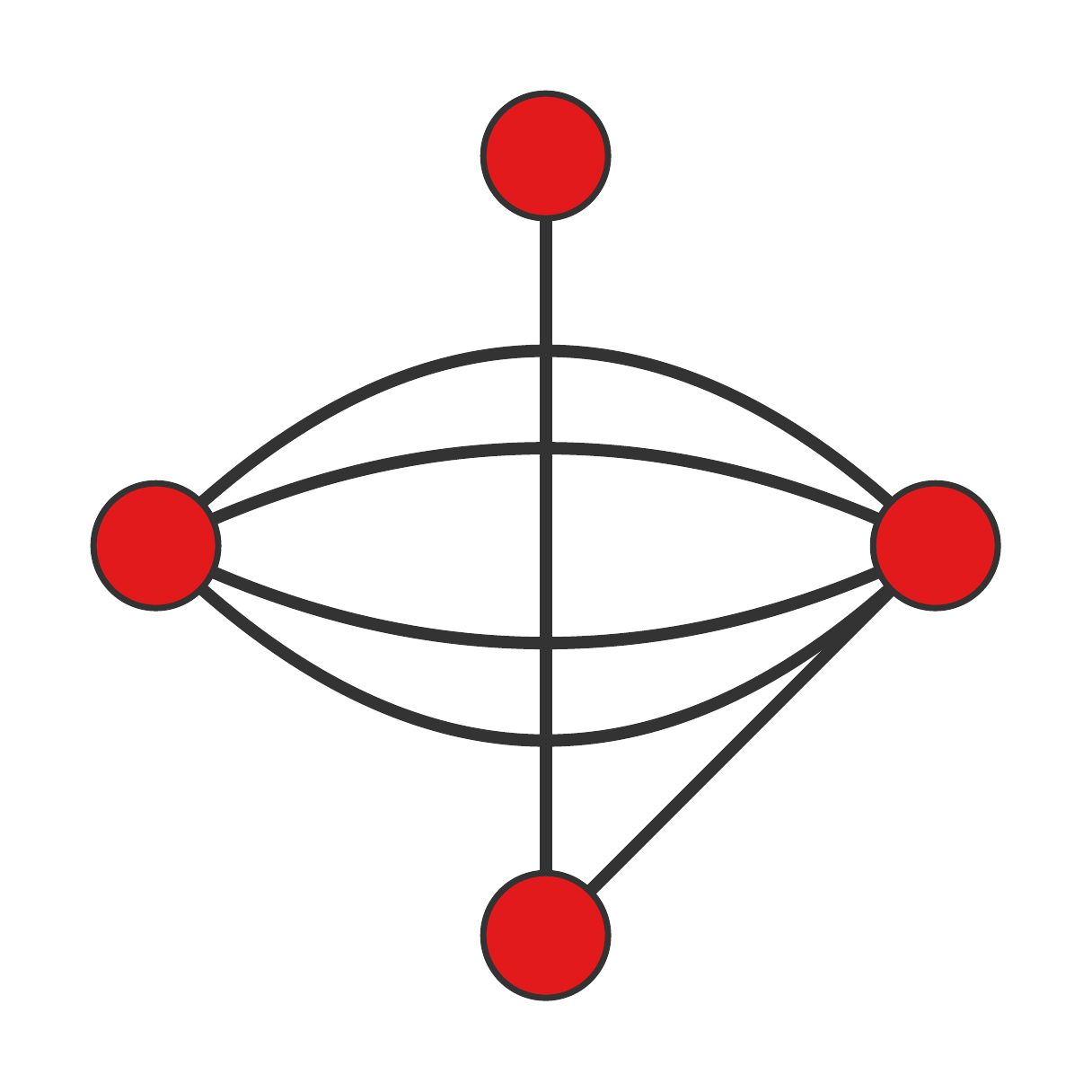}{1}{\frac{1}{4}}
    \end{tikzpicture}
	\caption{Distributions of the $\log_{10}$ of the selected EFPs as chosen by the black-box guided strategy, for prompt (signal) muons and non-prompt (background) muons.}
	\label{fig:efp_all}
\end{figure}

\subsection{IRC-unsafe Observables}

To understand the nature of the remaining information used by the PFN but not captured by the isolation cones and the IRC-safe observables, we expand the search space to include observables which are not IRC safe ($\kappa \in [-1,0,\frac{1}{4},\frac{1}{2},1,2]$), with alternative angular powers ($\beta \in [\frac{1}{4},\frac{1}{2},1,2,3,4]$) and with up to $n=7$ nodes and $d=7$ edges.

A scan of these observables finds a set of \rev{5} which, when combined with the isolation cones and $\sum p_{\textrm{T}}$ reach an AUC of \rev{0.857}. \rev{Figure~\ref{fig:efp_all} shows the EFP graphs as well as their distributions for prompt and non-prompt muons. They include  single point-graphs, with no angular powers, as well as a two-point correlators with large angular power sensitive to high-angle effects, and more complex graphs with multiple nodes.}  We note that due to the overlapping nature of the large space of EFPs, there are several sets of  EFPs which achieve similar performance.  Again, a similar scan guided by the CNN rather than the PFN yields very similar results.

\begin{table}[]
    \centering
        \caption{ Summary of  performance (AUC) in the prompt muon classification task for various network architectures and input features. Statistical uncertainty in each case is $\pm 0.001$ with 95\% confidence, measured using bootstrapping over 100 models. Uncertainty due to the initial conditions of the network is found to be negligible. Also shown are the number of \rvvv{inputs to and} parameters \rvvv{of} each network.}
        \label{tab:summary}
        \rev{
    \begin{tabular}{l|c|c|c|c}
    \hline\hline
    Method &  \rvvv{$N_\textrm{inputs}$} & AUC & ADO & $N_\textrm{Params}$\\
         &   &  &[PFN] & \\

    \hline
       Single Iso Cone & \rvvv{1} & {\ISOOneresult} & 0.860 & \rvv{40k}\\
       10 Iso   & \rvvv{10} & {\ISOTenresult} & 0.877& \rvv{41k}\\
       \rvv{ 10 Iso, $\sum p_{\textrm{T}}$ }  & \rvvv{11} & \rvv{0.807} & \rvv{0.884}& \rvv{42k}\\

       \rev{10} Iso, $\sum p_{\textrm{T}}$, \rev{4} simple EFPs & \rvvv{15} & \rvv{0.842} & \rvv{0.888} & \rvv{42k}\\
       \rev{10} Iso, $\sum p_{\textrm{T}}$, \rev{5} EFPs & \rvvv{16} & {0.857} & \rvv{0.900} & \rvv{43k}\\ 
       Calo image CNN & \rvvv{1024} & {\CNNresult}  &  \rvv{0.950} & 167k\\
       Calo cell Energy-Flow Net & \rvvv{102} & {0.849} & 0.951 & 453k\\
       Calo cell Particle-Flow Net & \rvvv{102} &{\PFNresult} & 1 & 453k\\
       \hline\hline
    \end{tabular}}
    \label{tab:my_label}
\end{table}

\section{Discussion}

The performance of the networks which use the low-level calorimeter cells indicates that information exists in these cells which is not captured by the isolation cones, see Table~\ref{tab:summary}.  A guided search through the space of \rev{IRC-safe} EFPs closes most of the gap between these networks, giving us some insight as to the nature of the information.  A broader search is able to complete the bridge, yielding the same performance as the low-level network, but employing IRC-unsafe EFPs.  \rvvv{The multi-point correlators may be sensitive to the width of the jet, due to the momentum of the constituents relative to the jet axis, as a result of $b$- and $c$-quark decays.} 

\rev{A comparison of the network complexity for the various approaches is shown in Tab.~\ref{tab:summary}.  The set of high-level features (isolation cones and EFP graphs)  matches the PFN performance with \rvv{10 times} fewer parameters, supporting the notion that the high-level features are effectively summarizing the relevant low-level information.}


\section{Conclusions}

We have applied deep networks to low-level calorimeter deposits surrounding prompt and non-prompt muons in order to estimate the amount of classification power available and to probe whether the standard methods are fully capturing the relevant information.

The performance of the calorimeter cell networks significantly exceeds the benchmark approach, a single isolation cone.  The use of several isolation cones provides some improvement, suggesting that there is additional useful information in the full radial energy distribution. However, a substantial gap remains, hinting \rev{that} there is non-radial structure in the calorimeter cells which provides useful information for classification.  We map the strategy of the calorimeter cell networks into a set of energy flow polynomials, finding four IRC-safe, simple three-point correlators which capture a significant amount of the missing information.  As they are simple functions of the energy deposition, they can be physically interpreted, and the fidelity of their modeling can be \rvv{studied in} control regions in collider data. \rvv{Any boost in the efficiency to identify prompt muons} is extremely valuable to searches at the LHC, especially those with multiple leptons, where event-level efficiencies depend sensitively on object-level efficiencies.

Additional, more complex EFPs provide a further modest boost in performance,  \rev{closing the gap with the PFN}. The strong performance of the IRC-safe EFN suggests that most of the additional information beyond the isolation cones is IRC-safe.

More broadly, the existence of a gap between the performance of state-of-the-art high-level features and networks using lower-level calorimeter information represents an opportunity to gather additional power in the battle to suppress lepton backgrounds. Rather than employing black-box deep networks directly, we have demonstrated the power of using them to identify the relevant observables from a large list of physically interpretable options.  This allows the physicist to understand the nature of the information being used and to assess its systematic uncertainty. \rvvv{Here we have focused on two-dimensional projections of the calorimeter response, but longitudinal information expressed in three dimensions may offer additional power in future work}.  While these studies were performed with simulated samples, similar studies can be performed using unsupervised methods~\cite{Dery:2017fap,Metodiev:2017vrx} on samples of collider data, which we leave to future studies.

\section{IX. Acknowledgements}

We would like to thank Michael Fenton, Dan Guest and Jesse Thaler for providing valuable feedback and insightful comments and Yuzo Kanomata for computing support. We also wish to acknowledge a hardware grant from NVIDIA. This material is based upon work supported by the National Science Foundation under grant number 1633631. DW is supported by the DOE Office of Science. The work of JC and PB in part supported by grants NSF 1839429 and NSF 
NRT 1633631 to PB.

\bibliography{muon}

\appendix

\section{Appendices}
\subsection{A. Neural Network Architectures}
\label{sec:nn}

All networks were trained in Tensorflow\cite{tensorflow2015} and Keras\cite{chollet2015keras}. 
The networks were optimized with Adam~\cite{Adam2014} for up to 100 epochs with early stopping.
For all networks except the PFNs, the weights were initialized using orthogonal weights\cite{OrthogonalWeights}.
Hyperparameters were optimized using Bayesian optimization with the Sherpa hyperparameter optimization library \cite{SherpaOptimization}. The variables and ranges for the hyperparameters are shown in tables \ref{tab:hyperparam_range_conv} and \ref{tab:hyperparam_range}.

Below are further details regarding the networks which use images and those which use isolation and EFP observables.

\subsubsection{B. Muon Image Networks}

The pixelated images were preprocessed to have zero mean and unit standard deviation. \rev{We tried rotating the images as in \cite{Baldi:2016fql} but performance was considerably lowered by this preprocessing step.} The best muon image network structure begins with three convolutional blocks. Each block contains \rvv{three} convolutional layers
with \rvv{48} filters with rectified linear units~\cite{RectifiedLinearUnits}, followed by a 2x2 pooling layer. 
Afterwards there are \rvv{two} fully connected layers with \rvv{74} rectified linear units and a final layer with a sigmoidal logistic activation function to classify signal vs background.
The model had dropout~\cite{srivastava2014dropout,baldidropout14} with value \rvv{0.2388} on the fully connected layers and an initial learning rate of \rvv{0.0003} and batch size of 128.

\begin{table}[h]
    \centering
        \caption{Hyperparameter ranges for bayesian optimization of convolutional networks}
    \begin{tabular}{c|c|c}
    \hline\hline
        Parameter &  Range & Value\\
        \hline
          Num. of convolutional blocks & [1, \rvv{4}] & {3} \\
          Num. of filters & [16, 128] & {48}  \\
          Num. of fully connected layers & [2, \rvv{4}] & {2} \\
          Number of hidden units & [25, 200] & {74}\\
          Learning rate & [0.0001, 0.01] & {0.0003} \\
          Dropout & [0.0, 0.5] & {0.2388} \\
             \hline\hline
    \end{tabular}
    \label{tab:hyperparam_range_conv}
\end{table}

\subsubsection{C. Particle-Flow Networks}

The Particle Flow Network (PFN) is trained using the \verb@energyflow@ package\cite{Komiske_2019}. Input features are taken from the muon image pixels and preprocessed by subtracting the mean and dividing by the variance. The PFN uses 3 dense layers in the per-particle frontend module and 3 dense layers in the backend module. Each layer uses 100 nodes, \verb@relu@ activation and \verb@glorot_normal@ initializer. The final output layer uses a sigmoidal logistic activation function to predict the probability of signal or   background. The \verb@Adam@ optimizer is used with a learning rate of 0.0001 and trained with a batch size of 128. 

\subsubsection{D. Isolation Cone \rvv{and EFP} Networks}

The isolation inputs \rvv{and EFPs} are preprocessed by subracting the mean and dividing by the variance. We trained neural networks with two to eight fully connected hidden layers depending on the hyperparameter value and a final layer with a sigmoidal logistic activation function to predict the probability of signal or background. 

For the minimal set of isolation inputs, the best model we found had \rvv{2} fully connected layers with \rvv{197} rectified linear hidden units\cite{RectifiedLinearUnits} and a learning rate of \rvv{0.0003} and dropout rate of \rvv{0.0547}.

\begin{table}[h]
    \centering
        \caption{Hyperparameter ranges for Bayesian optimization of fully connected networks
        }
    \begin{tabular}{c|c|c}
    \hline\hline
        Parameter &  Range & ISO Value 
        \\
        \hline
          Num. of layers & [2, 8] & {2} 
          \\
          Num. of hidden units & [1, 200] & {197} 
          \\
          Learning rate & [0.0001, 0.01] & {0.0003}
          \\
          Dropout & [0.0, 0.5] & {0.0547} 
          \\
             \hline\hline
    \end{tabular}
    \label{tab:hyperparam_range}
\end{table}

\subsection{ADO comparison}

\begin{figure}
    \centering
    \includegraphics[width=0.45\textwidth]{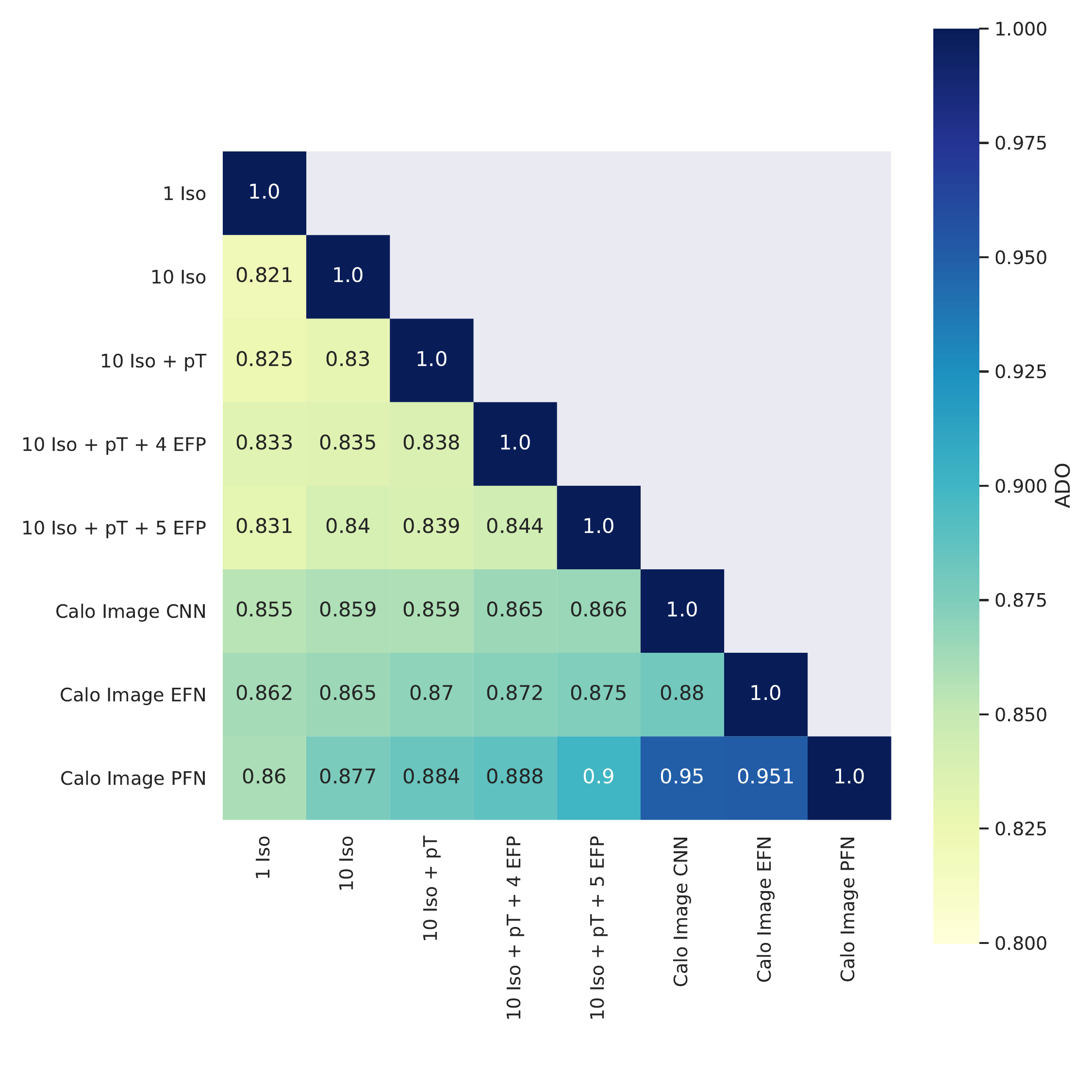}
    \caption{ Comparison of the similarity of decisions made by pairs of networks, as quantified by the Average Decision Ordering (ADO)~\cite{Faucett:2020vbu}, defined in the text.}
    \label{fig:ado_comp}
\end{figure}

In Fig.~\ref{fig:ado_comp}, the ADO between the various networks is shown.

\end{document}